\documentclass[draftclsnofoot,onecolumn, 12pt]{IEEEtran}
\usepackage{graphicx}
\usepackage{subfigure}
\usepackage{epstopdf}
\usepackage{graphicx}
\usepackage{amsmath}
\usepackage{amsfonts}
\usepackage{amssymb, color, bm}
\usepackage{bbm}

\usepackage{algorithmicx}
\usepackage[noend]{algpseudocode}
\usepackage{algorithm,setspace}
\usepackage{tabularx}
\usepackage[T1]{fontenc}

\usepackage{multicol}
\usepackage{lipsum}
\usepackage{mwe}

\DeclareMathOperator{\E}{\mathbb{E}}
\DeclareMathOperator{\I}{\mathbb{I}}

\usepackage[utf8]{inputenc}
\allowdisplaybreaks
\usepackage{cite}

\DeclareMathOperator{\R}{\mathbb{R}}

\usepackage{lineno}

\title{{\color{black}{Modular Meta-Learning for Power Control via Random Edge Graph Neural Networks}}}
\author{Ivana Nikoloska,~\IEEEmembership{Graduate Student Member,~IEEE,}
        and~Osvaldo Simeone,~\IEEEmembership{Fellow,~IEEE.}
\thanks{I. Nikoloska and O. Simeone are with KCLIP, CTR, Dept. of Engineering, King's College London e-mails: \{ivana.nikoloska, osvaldo.simeone\}@kcl.ac.uk.}
\thanks{This work was supported by the European Research Council (ERC) under the European Union’s Horizon 2020 Research and Innovation Program (Grant Agreement No. 725731).}%
}

\begin{document}

\maketitle

\begin{abstract}
    In this paper, we consider the problem of power control for a wireless network with an arbitrarily time-varying topology, including the possible addition or removal of nodes. A data-driven design methodology that leverages graph neural networks (GNNs) is adopted in order to efficiently parametrize the power control policy mapping the channel state information (CSI) to transmit powers. The specific GNN architecture, known as random edge GNN (REGNN), defines a non-linear graph convolutional filter whose spatial weights are tied to the channel coefficients. While prior work assumed a joint training approach whereby the REGNN-based policy is shared across all topologies, this paper targets adaptation of the power control policy based on limited CSI data regarding the current topology. {\color{black}{To this end, we propose a novel modular meta-learning technique that enables the efficient optimization of module assignment. While  black-box meta-learning optimizes a general-purpose adaptation procedure via (stochastic) gradient descent, modular meta-learning finds a set of reusable modules that can form components of a solution for any new network topology.}}  Numerical results validate the benefits of meta-learning for power control problems over joint training schemes, and demonstrate the advantages of modular meta-learning when data availability is extremely limited.
\end{abstract}


\section{Introduction}

\subsection{Motivation}

With the proliferation of wireless devices and
services, wireless communication networks are becoming increasingly complex. Beyond 5G (B5G) networks are expected to provide uninterrupted connectivity to devices ranging from sensors and cell phones to vehicles and robots, calling for the development of novel interference management strategies via radio resource management (RRM). However, solving most RRM problems is NP-hard, making it challenging to derive an optimal solution in all but the simplest scenarios \cite{mollanoori2013uplink}.

Solutions to this problem run the gamut from classical optimization techniques \cite{lei2015joint} to information and game theory \cite{yang2017mean,riaz2018power}. As emerging applications demand growth in scale and complexity, modern machine learning techniques have also been explored as alternatives to solve RRM problems in the presence of model and/or algorithmic deficits \cite{simeone2018very}.
The performance of trained models generally depend on how representative the training data are for the channel conditions encountered at deployment time. As a result, when conditions in the network change, these rigid models are often no longer useful \cite{nair2019covariate}, \cite{quinonero2009dataset}.

A fundamental RRM problem is the optimization of transmission power levels at distributed links that share the same spectral resources in the presence of time-varying channel conditions \cite{chiang2008power}.
This problem was addressed by the data-driven methodology introduced in \cite{eisen2020optimal}, and later studied in \cite{eisen2020transferable,naderializadeh2020wireless,chowdhury2020unfolding}. In it, the power control policy mapping channel state information (CSI) and power vector is parametrized by a graph neural network (GNN). The GNN encodes information about the network topology through its underlying graph whose edge weights are tied to the channel realizations. The design problem consists of training the weights of the graph filters, while tying the spatial weights applied by the GNN to the CSI. As a result, the solution -- which is referred to as random edge GNN (REGNN) -- automatically adapts to time-varying CSI conditions.

{\color{black}{To improve data and iteration efficiency, in this paper, we focus on the higher-level problem of facilitating adaptation to time-varying topologies,  allowing also for a variable number of nodes over time.}} To this end, as illustrated in Fig.~\ref{sys_mod}, we assume that the topology of the network varies across periods of operation of the system, with each period being characterized by time-varying channel conditions as in \cite{eisen2020optimal}. As such, the operation within each channel period is well reflected by the model studied in \cite{eisen2020optimal,naderializadeh2020wireless}, and we adopt an REGNN architecture for within-period adaptation. At the beginning of each period, the network designer is given limited CSI data that can be used to adapt the REGNN-based power control policy to the changed topology. In order to facilitate fast adaptation -- in terms of data and iteration requirements -- we integrate \textit{meta-learning} with REGNN training.

\subsection{Meta-learning}
{\color{black}{The goal of meta-learning is to extract shared knowledge, in the form of an inductive bias, from data sets corresponding to distinct learning tasks in order to solve held-out tasks and adapt to new network topologies more efficiently \cite{schmidhuber1987evolutionary, thrun1998lifelong}.}} The inductive bias may refer to parameters of a general-purpose learning procedure, such as the learning rate \cite{maclaurin2015gradient}, or initialization \cite{finn2017model}, \cite{nichol2018firstorder}, \cite{grant2018recasting} of (stochastic) gradient descent (S)GD. These schemes can be credited for much of the reinvigorated interest in meta-learning in the previous decade. We will refer to them as \textit{black-box meta-learning} methods, given their model-agnostic applicability via fast parametric generalization.

In contrast, \textit{modular meta-learning} aims at fast combinatorial generalization \cite{chomsky2014aspects}, making, in a sense, "infinite use of finite means" \cite{von1999humboldt}. Modular meta-learning generalizes to new tasks by optimizing a set of neural network modules that can be composed in different ways to solve a new task, without changing their internal parameters \cite{alet2018modular}, \cite{alet2019neural}. Modularity is a key property of engineered systems, due to its fault tolerance, interpretability, and flexibility \cite{baldwin2006modularity}, but is generally lacking in data-driven solutions, which often amount to large black-box input-output mappings. 
The few existing modular meta-learning approaches rely on simulated annealing to find a suitable module composition for each task given the current neural network modules \cite{alet2018modular}. These, however, are notoriously inefficient optimization methods (in terms of computation time), and more recent techniques integrate learnt proposal functions in order to speed up training \cite{alet2019neural}.

\subsection{Contributions}
{\color{black}{As illustrated in Fig.~\ref{sys_mod}, the main goal of this paper is to optimize fast adaptation procedures for the power control policy that are able to cope with time-varying network configurations. Fast adaptation cannot be accomplished by existing optimization and learning methods that search for a new optimized solution for each new network topology. To address this limitation in the state of the art, we introduce the use of meta-learning for the problem of power control in time-varying networks by studying both black-box and modular meta-learning methods.}} {\color{black}{Specifically, the main contributions of this paper can be summarized as follows:

\begin{itemize}
    \item We introduce a novel modular meta-learning method that constructs a repository of fixed graph filters that can be combined to define REGNN-based power control models for new network configurations. In contrast to existing modular meta-learning schemes that rely on variants of global optimization via simulated annealing, the proposed method adopts an efficient stochastic module assignment based on the Gumbel-softmax reparametrization trick \cite{maddison2016concrete}, which enables efficient optimization via standard SGD at run time.

    \item To highlight conditions under which modular meta-learning may be  beneficial over the better established black-box meta-learning methods, we develop a solution that integrates first-order model agnostic meta-learning (FOMAML) \cite{finn2017model}, a state-of-the-art representative of black-box meta-learning methods, with REGNN training.
    
    \item We validate the performance of all meta-learning methods with extensive experiments that provide comparisons with joint training schemes \cite{eisen2020optimal}. The use of meta-learning for power control problems in wireless networks is validated, and a comparative study of the performance of the considered meta-learning solutions is presented.
\end{itemize}}}

\subsection{Prior Work}
{\color{black}{ Power control is one of the oldest and most important problems in communication engineering, and has received significant attention from the communication community. Different methods and techniques have been proposed, with the most recent ones relying on novel deep learning techniques \cite{sun2018learning}, \cite{liang2019towards}. Whilst very promising, such methods rely on the use of fully-connected deep learning models, with input and output layers of fixed sizes. Therefore, these techniques are not applicable to the settings of interest in this study, in which the number of nodes  varies over time.}}

Learning with inputs and outputs of variable size can be done using geometric models, such as GNNs. In fact, GNNs are enjoying an increasing popularity in the wireless communication community. In addition to power allocation \cite{eisen2020optimal,eisen2020transferable,naderializadeh2020wireless,chowdhury2020unfolding}, GNNs have been used to address cellular \cite{zhao2020cellular} and satellite \cite{yang2020noval} traffic prediction, link scheduling \cite{lee2020graph}, channel control \cite{tekbiyik2020channel}, and localization \cite{yan2021graph}. Due to their localized nature, GNNs have also been applied to cooperative \cite{dong2020drl} and decentralized \cite{lee2021decentralized} control problems in networked systems. A review of the use of GNNs in wireless communication can be found in \cite{he2021overview}.  

Meta-learning has been shown to improve the training and adaptation efficiency in various problems in wireless communications, ranging from demodulation \cite{park2020learning} and decoding \cite{jiang2019mind}, to channel estimation \cite{mao2019roemnet} and beamforming \cite{yuan2020transfer}. In particular, in \cite{park2020learning} the authors use pilots from previous transmissions of Internet of Things (IoT)
devices in order to adapt a demodulator to new channel conditions using few pilot symbols. The authors of \cite{mao2019roemnet}  train a neural network-based channel estimator for orthogonal frequency-division multiplexing (OFDM) system with FOMAML in order to obtain an effective solution given a small number of samples. Reference \cite{yuan2020transfer} studies fast beamforming in multiuser multiple-input single-output (MISO) downlink systems. 
An overview of meta-learning methods, with applications to wireless communication networks is available in \cite{simeone2020learning}.

The application of meta-learning to GNN-based power control was presented in the conference version of this paper for the first time \cite{nikoloska2021fast}. In particular, \cite{nikoloska2021fast} considers black-box methods and offers preliminary experimental results. In contrast to the preliminary conference version \cite{nikoloska2021fast}, in this paper, we consider both black-box and modular meta-learning solutions, and we provide a more comprehensive numerical evaluation of all considered meta-learning schemes. To the best of the authors' knowledge, this is the first work investigating the use of modular meta-learning in communication engineering problems. 

The rest of the paper is organized as follows. The considered model and problem are presented in Section~\ref{sec_mod_prob}, and REGNNs are reviewed in Section~\ref{sec_regnn}. Meta-learning is introduced in Section~\ref{meta_over}, and black-box methods and the proposed modular solution are given in Section~\ref{sec_meta} and Section~\ref{sec_meta_mod}, respectively. 
All meta-learning schemes are evaluated in Section~\ref{sec_num}. Section~\ref{sec_con} concludes the paper.

\section{Model and Problem}\label{sec_mod_prob}
\begin{figure*}[tbp]
\centering
\includegraphics[width=0.86\linewidth]{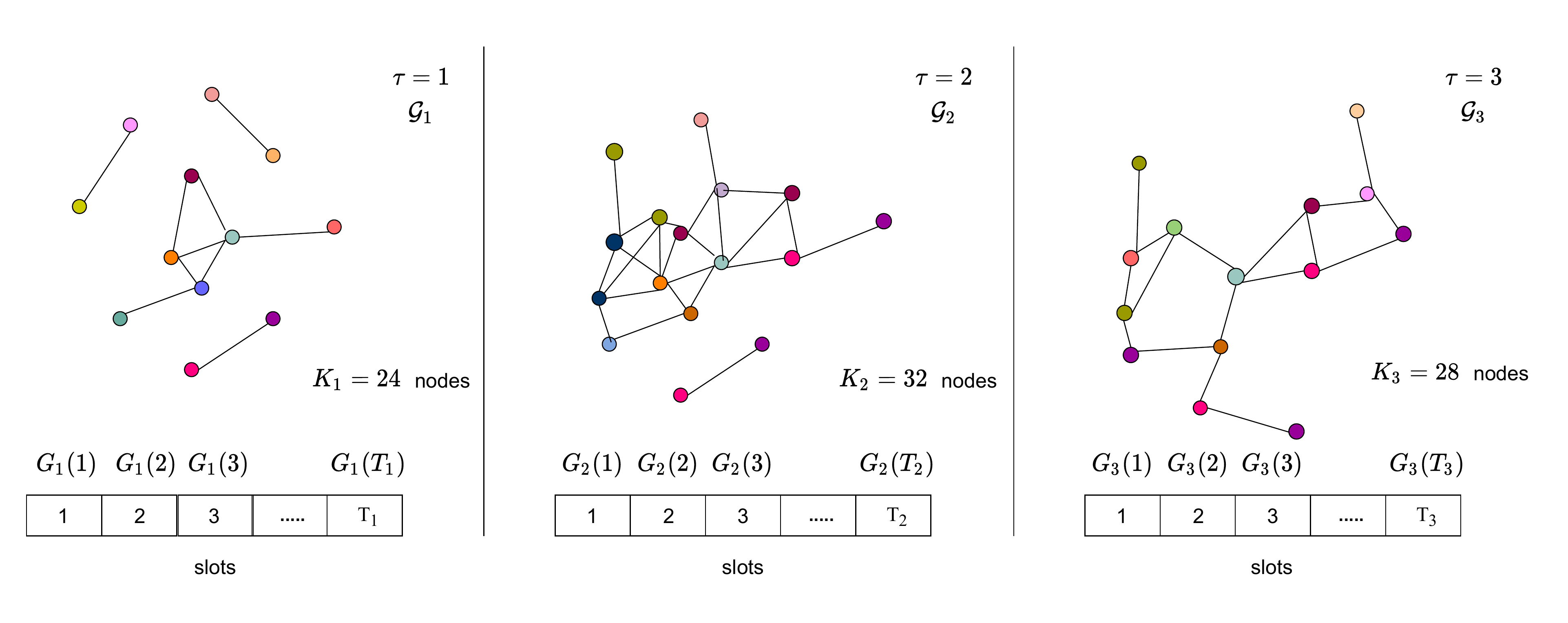}
\vspace*{-3mm}
\caption{Interference graph $\mathcal{G}_\tau$ over periods $\tau = 1$, $\tau = 2$, and $\tau = 3$. Each vertex represents a communication link, and an edge is included between interfering links. }
\label{sys_mod}
\end{figure*}

As illustrated in Fig.~\ref{sys_mod}, we consider a wireless network running over periods $\tau=1,...,\mathcal{T}$, with topology possibly changing at each period $\tau$. During period $\tau$, the network is comprised of $K_\tau$ communication links. Transmissions on the $K_\tau$ links are assumed to occur at the same time using the same frequency band.  The resulting interference graph $\mathcal{G}_\tau = (\mathcal{K}_\tau, \mathcal{E}_\tau)$ includes an edge $(k, j) \in \mathcal{E}_\tau$ for any pair of links $k, j \in \mathcal{K}_\tau$ with $k \neq j$ whose transmissions interfere with one another. We denote by $\mathcal{N}^k_\tau \subseteq \mathcal{K}_\tau$ the subset of links that interfere with link $k$ at period $\tau$. Both the number of links $K_\tau = |\mathcal{K}_\tau|$ and the topology defined by the edge set $\mathcal{E}_\tau$ generally vary across periods $\tau$. 

Each period contains $T_\tau$ time slots, indexed by $t = 1,...,T_\tau$. In time slot $t$ of period $\tau$, the channel between the transmitter of link $k$ and its intended receiver is denoted by $h^{k,k}_\tau (t)$, while $h^{j,k}_\tau (t)$ denotes the channel between transmitter of link $j$ and receiver of link $k$ with $j \in \mathcal{N}^k_\tau$. Channels account for both slow and fast fading effects, and, by definition of the interference graph $\mathcal{G}_\tau$, we have $h^{j,k}_\tau (t) = 0$ for $j \notin \mathcal{N}^k_\tau$. The channels for slot $t$ in period $\tau$ are arranged in the channel matrix $G_\tau (t) \in \R^{K_\tau \times K_\tau}$, with the $(j,k)$ entry given by $\left[ G_\tau(t) \right]_{j,k} = g^{j,k}_\tau (t) = |h^{j,k}_\tau (t)|^2$. 
Channel states vary across time slots, and the marginal distribution of matrix $G_\tau(t)$ for all $t = 1,...,T_\tau$ is constant and denoted by $\mathcal{P}_\tau (G_\tau)$. The distribution $\mathcal{P}_\tau (G_\tau)$ generally changes across periods $\tau$, and it is a priori unknown to the network.

To manage inter-link interference, it is useful to adjust the transmit powers such that a global network-wide objective function is optimized (see, e.g., \cite{douros2011review}). For each channel realization $G_\tau (t)$, we denote the vector of power allocation variables as $p_\tau (t) \in \R^{K_\tau}$, whose $k$-th component, $\left[ p_\tau(t) \right]_{k} = p^k_\tau (t)$, represents the per-symbol transmit power of transmitter $k$ at time slot $t$ of period $\tau$. The resulting achievable rate in bits per channel use for link $k$ is given by
\begin{align}\label{rate_k}
    c^k(G_\tau (t), p_\tau (t)) = \log_2 \left(1 + \frac{g^{k,k}_\tau (t) p^k_\tau (t)}{\sigma^2 + \sum_{j \in \mathcal{N}_\tau^k} g^{j,k}_\tau(t) p^j_\tau(t)}\right),
\end{align}
where $\sigma^2$ denotes the per-symbol noise power. By \eqref{rate_k}, interference is treated as worst-case additive Gaussian noise.

The goal of the system is to determine a power allocation policy $\textrm{p}_\tau(\cdot)$ in each period $\tau$ that maps the channel matrix $G_\tau(t)$ to a power allocation vector $p_\tau (t)$ as
\begin{align}\label{power_f}
    p_\tau (t) = \textrm{p}^k_\tau(G_\tau (t))
\end{align}
by maximizing the average achievable sum-rate. This yields the stochastic optimization problem
\begin{align}\label{opt}
    &\underset{\textrm{p}_\tau(\cdot)}{\text{max}} \,\,\,\, \sum_{k = 1}^{K} \E_{G_\tau \sim \mathcal{P} (G_\tau)} \Big[c^k(G_\tau, p_\tau(G_\tau))\Big] \nonumber\\
    & \text{s.t.} \,\,\,\,  0 \leq \textrm{p}^k_\tau(\cdot) \leq P^k_{\textrm{max}}, \,\,\,\, \text{for} \,\,\,\, k = 1,...,K,
\end{align}
where $P^k_{\textrm{max}}$ denotes the power constraint of link $k$. Note that problem \eqref{opt} is defined separately for each period $\tau$. Since the distribution $\mathcal{P} (G_\tau)$ is unknown, problem \eqref{opt} can not be addressed directly. 

We assume, however, that the designer has access to channel realizations $\{G_\tau (1), ..., G_\tau (T_\tau)\}$ over $T_\tau$ time slots in period $\tau$. {\color{black}{The required channel information can be obtained using suitable acquisition techniques (e.g., \cite{burghal2015efficient}). Similarly to \cite{eisen2019large}, we do not consider the effect of imperfect channel information, whose investigation is left for future work.}} Accordingly, problem \eqref{opt} can be approximated by estimating the objective in \eqref{opt} via an empirical average as in
\begin{align}\label{opt_approx}
    &\underset{\textrm{p}_\tau(\cdot)}{\text{max}} \,\,\,\, \sum_{k = 1}^{K} \sum_{t = 1}^{T_\tau} c^k(G_\tau (t), p_\tau(G_\tau(t))) \nonumber\\
    & \text{s.t.} \,\,\,\,  0 \leq \textrm{p}^k_\tau(\cdot) \leq P^k_{\textrm{max}}, \,\,\,\, \text{for} \,\,\,\, k = 1,...,K.
\end{align}
{\color{black}{We are interested in cases where the number of transmitters is potentially large. This makes the solution of \eqref{opt_approx} intractable, and it motivates the use of approximately optimal methods, including data driven approaches.}}

\begin{figure}[tbp]
\centering
\includegraphics[width=0.9\linewidth]{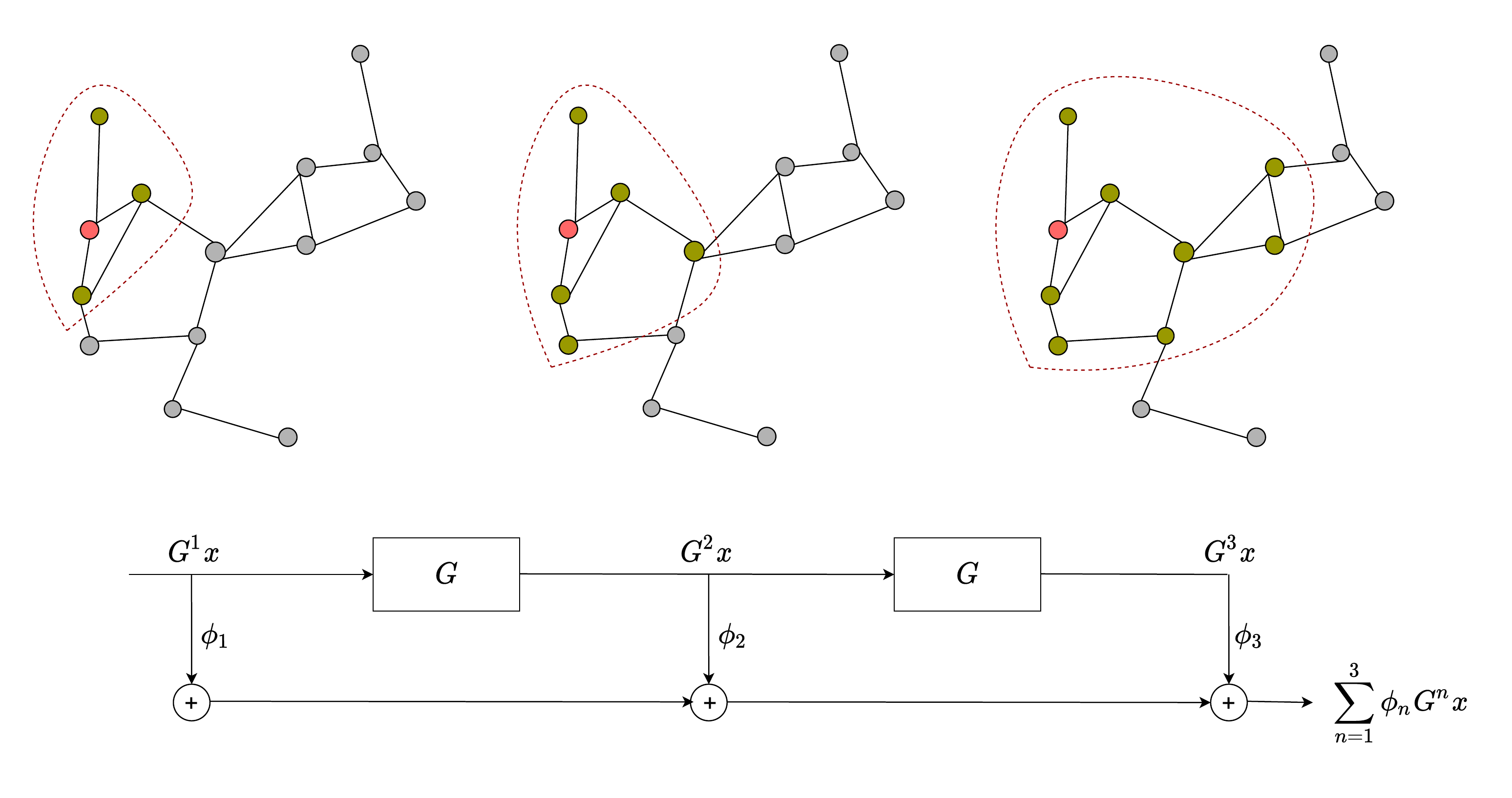}
\caption{A graph convolutional filter is a polynomial on a matrix representation of the interference graph. As the order $N$ of the filter grows (here $N=3$), information is aggregated from nodes that are farther apart (shown in green for the output corresponding to the node in red.)}
\label{regnn_model}
\end{figure}

\section{Power Allocation by Training REGNN}\label{sec_regnn}
In this section, we review the solution proposed in \cite{eisen2020optimal}, which tackles problem \eqref{opt_approx}
separately for each period $\tau$. 
The approach in \cite{eisen2020optimal} parametrizes the power allocation function $\textrm{p}_\tau(\cdot)$ in \eqref{power_f} by a REGNN as 
\begin{align}\label{power_param}
    p_\tau(G_\tau)  = \textrm{f} (G_\tau \, | \, \Phi_\tau),
\end{align}
where $\Phi_\tau \in \R^N$ is a vector of trainable parameters. In the rest of this section, we first describe the mapping $\textrm{f} (\cdot \, | \, \Phi_\tau)$ implemented by a REGNN, and then we review the problem of optimizing the parameter vector $\Phi_\tau$. Unless stated otherwise, in this section we drop the index $\tau$, which is fixed in order to simplify notation.

\subsection{REGNN Model}
To introduce the REGNN model, let us first describe the key operation of graph filtering. Consider a graph $\mathcal{G} = (\mathcal{N}, \mathcal{E})$, with $K$ nodes in set $\mathcal{N}$ and edge set $\mathcal{E}$. We associate to graph $\mathcal{G}$ a $K \times K$ matrix $G$, known as the graph shift operator (GSO), with the property that we have $\left[ G\right]_{j,k} = 0$ for $(j,k) \notin \mathcal{E}$. Note that the channel matrix satisfies this condition for the interference graph. A graph signal is a $K \times 1$ vector $x$, with each entry being assigned to one of the nodes in the graph.  Given a $N \times 1$ vector of filter taps $\phi = [\phi_1, ... , \phi_N]^T$ with $\phi_n \in \R$, a graph filter applies the graph convolution \cite{sandryhaila2013discrete}
\begin{align}\label{regnn}
    \phi \ast_G x = A(G) x = \sum_{n=1}^N \phi_{n} G^n x
\end{align}
to a $K \times 1$ input graph signal $x$.
The filter $A(G) = \sum_{n=1}^N \phi_{n} G^n$ is a polynomial of the matrix $G$. 

As illustrated in Fig.~\ref{regnn_model},
each $n$-th power of $G^n$ of the GSO \eqref{regnn} performs an $n$-hop shift $G^n x$ of the elements in vector $x$ on the graph. Specifically, the term $G x$ is a $K \times 1$ vector whose $k$-th entry aggregates the entries in vector $x$ corresponding to  single-hop neighbouring nodes $j \in \mathcal{N}^k$ of node $k$, each weighted by the corresponding channel element $\left[ G \right]_{j,k}$ of the GSO; the term $G^2 x$ aggregates for each node the contributions in vector $x$ associated to two-hop neighbouring nodes; and so on. As illustrated in Fig.~\ref{regnn_model}, as the order $n$ increases, node inputs from larger neighborhoods are incorporated. Thus, the graph convolution implements a local message-passing procedure, with information from larger neighbourhoods being aggregated as the filter size $N$ in \eqref{regnn} increases.

An REGNN consists of a layered architecture in which each layer is a composition of a graph convolution and a per-node non-linearity. The graph convolution in each layer uses the current channel matrix $G(t)$ as the GSO in \eqref{regnn}. Due to its dependence on the random fading channels, the graph convolution is characterized by "random edges" according to the terminology used in \cite{eisen2020optimal}. Given the current channel matrix $G(t)$, the output of each $l$-th intermediate layer is given as
\begin{align}\label{regnn1}
    z_{l+1} = \sigma \left[ \sum_{n=1}^N \Phi_{l,n} G(t)^n z_l \right],
\end{align}
where $\sigma [\cdot]$ denotes a non-linear function, such as a rectified linear unit (ReLU) or a sigmoid, that is applied separately to each of the $K$ entries in the input.
The REGNN is defined by the recursive application of \eqref{regnn1} for $L$ layers, with input to the first layer given by the $K \times 1$ input graph signal $z_0 = x$. In this paper, the input signal $x$ is set to an all-one vector \cite{naderializadeh2020wireless}, but it may more generally include a variable describing the state of each link \cite{eisen2020optimal}. 

The transmit power in \eqref{power_param} is found as the output of the final, $L$-th layer of the REGNN as
\begin{align}\label{regnn2}
    &\textrm{f} (G(t) \, | \, \Phi) = P_\text{\textrm{max}} 
    \times \sigma \left[ \sum_{n=1}^N \Phi_{L,n} G(t)^n \left( ... \left(\sigma \left[ \sum_{n=1}^N \Phi_{1,n} G(t)^n x \right]\right) ...\right) \right],
\end{align}
with $P_\text{\textrm{max}}$ being a diagonal matrix with its $k$-th element on the main diagonal being given by $ \left[ P^k_{\textrm{max}} \right]_{k,k} = P^k_{\textrm{max}}$, and $\Phi_l = [\Phi_{l,1},...,\Phi_{l,N}]^T$, denoting the model parameters (convolution taps) for layer $l$. By \eqref{regnn2}, specifying the REGNN architecture requires defining the number of layers and the number of filter taps per layer. Assuming all layers have an equal number of taps, the total number of trainable parameters is thus $LN$, a number considerably smaller than what would be required to train a fully-connected neural network.

\subsection{Training a REGNN}
Given a set of channel realizations $\{G_\tau (1), ..., G_\tau (T)\}$ for a given period $\tau$, training of the REGNN parameters $\Phi_\tau = [\Phi_{1},...,\Phi_{L}] \in \R^{N \times L}$ is done by tackling the unsupervised learning problem \cite{eisen2020optimal}
\begin{align}\label{opt_phi}
    &\underset{\Phi}{\text{max}} \,\,\,\, \sum_{k = 1}^{K} \sum_{t = 1}^{T}  c^k(G_\tau (t), \textrm{f} (G_\tau (t) \, | \, \Phi)),
\end{align}
via (S)GD. Note that problem \eqref{opt_phi} restricts the optimization in \eqref{opt_approx} to the class of REGNNs in \eqref{regnn2}. By incorporating the channel matrices $G_{\tau}(t)$ in the structure of the REGNN-based power control policy $\textrm{p}(G_{\tau}(t)) = \textrm{f} (G_\tau (t) \, | \, \Phi_\tau)$, the method proposed in \cite{eisen2020optimal} automatically adapts to the different per-slot channel realizations.

\section{Meta-learning Power Control}\label{meta_over}

Our main goal in this paper is to improve the data efficiency of the REGNN solution reviewed in the previous section by enabling the explicit  adaptation of the power control policy $\textrm{p}_\tau(\cdot)$ to the interference graph of each period $\tau$, and hence across the changing topologies (see Fig.~\ref{sys_mod}). {\color{black}{Instead of learning a new policy for each new channel topology, which demands the availability of large channel information datasets, we propose to transfer knowledge across a number of previously observed topologies in the form of an adaptation procedure for the power control policy. This is done by meta-learning.}}

In order to enable meta-learning, we assume the availability of channel information from $\mathcal{T}^{\text{meta}}$ previous periods.
We denote the meta-training data set as $\mathcal{D} = \{\mathcal{D}_\tau\}_{\tau = 1,...,\mathcal{T}^{\text{meta}}}$, with $\mathcal{D}_\tau = \{G_\tau (1), ..., G_\tau (T_\tau)\}$ being the $T_\tau$ channel matrices available for each period $\tau$.
Following standard practice in meta-learning, each meta-training data set $\mathcal{D}_\tau$ is split into training data $\mathcal{D}^{\text{tr}}_\tau$ and testing data $\mathcal{D}^{\text{te}}_\tau$ \cite{finn2017model}, \cite{simeone2020learning}, and we write $t \in \mathcal{D}^{\text{tr}}_\tau$ and $t \in \mathcal{D}^{\text{te}}_\tau$ to denote the indices of the slots assigned to each set. 
At test time, during deployment, the network observes a new topology $\mathcal{G}_{\tau_\text{test}}$ for which it has access to a data set $\mathcal{D}^{\text{tr}}_{\tau_\text{test}}$, which is generally small, to optimize the power allocation strategy. 

The idea underlining meta-learning is to leverage the historical data $\mathcal{D}$ in order to optimize a learning algorithm $\Phi_\tau = \mathcal{A}(\mathcal{D}^{\text{tr}}_\tau)$ that uses training data $\mathcal{D}^{\text{tr}}_\tau$ to obtain a well performing REGNN parameter vector $\Phi_\tau$ for any new period $\tau$, even when the training data set $\mathcal{D}^{\text{tr}}_\tau$ is of limited size. In practice, the training algorithm $\mathcal{A}(\cdot)$ is either explicitly or implicitly defined by the solution of the learning problem \eqref{opt_phi} using the training data $\mathcal{D}^{\text{tr}}_\tau$. 
The meta-training objective is represented as the optimization problem
\begin{align}\label{meta_general}
    \underset{\mathcal{A}(\cdot)}{\text{max}} \,\,\,\, & \sum_{\tau = 1}^{\mathcal{T}^{\text{meta}}} \sum_{k = 1}^{K_\tau} \sum_{t \in \mathcal{D}^{\text{te}}_\tau} c^k(G_\tau (t), \textrm{f} (G_\tau (t) \, | \, \Phi_\tau = \mathcal{A}(\mathcal{D}^{\text{tr}}_\tau))),
\end{align}
where the testing part of the per-period data set $\mathcal{D}^{\text{te}}_\tau$ is used to obtain an unbiased estimate of the sum-rate in \eqref{opt}.

In the next two sections, we describe two approaches to formulate and solve the meta-learning problem \eqref{meta_general}. First, we adapt black-box meta-learning strategies that are based on a model-agnostic optimization approach \cite{finn2017model},\cite{nichol2018firstorder}. Then, we introduce a novel modular meta-learning method, which aims at discovering common structural elements for the power allocation strategies across different interference graphs. 

\begin{figure}[tbp]
\centering
\includegraphics[width=0.9\linewidth]{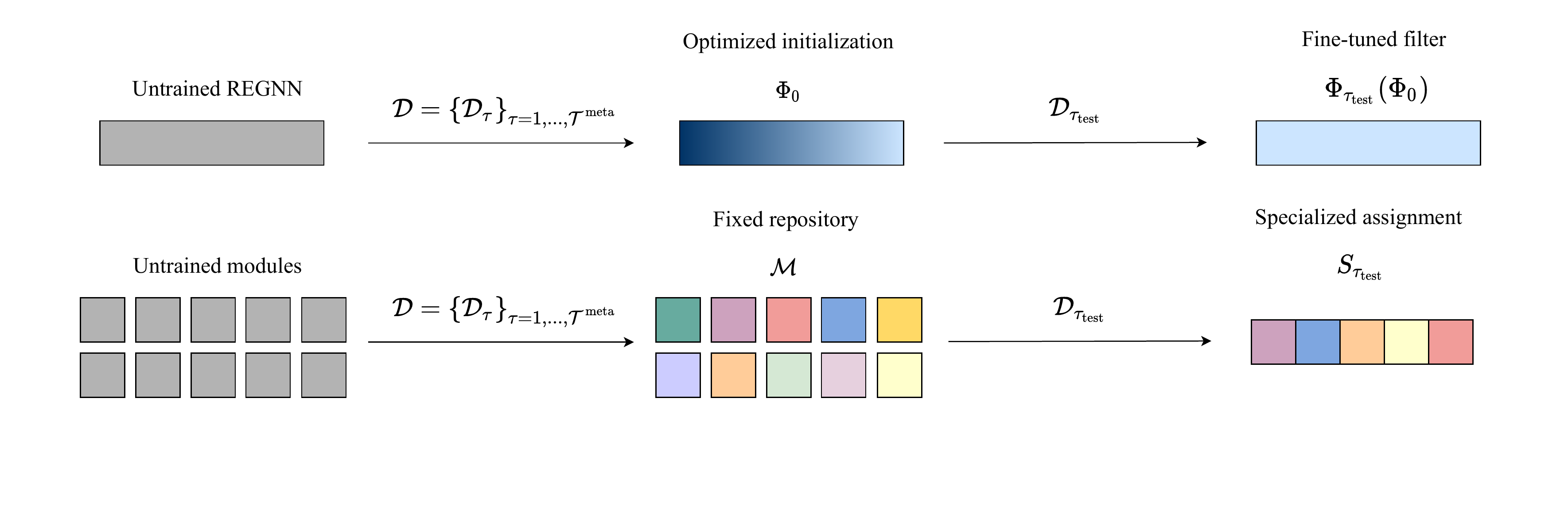}
\caption{(Top) Black-box strategies such as FOMAML optimize a representation that can be quickly adjusted to solve a new task. (Bottom) Modular meta-learning methods optimize a repertoire of
modules that can be quickly recombined at runtime to solve a new task.}
\label{ML_comparison}
\end{figure}

\section{Black-box Meta-learning}\label{sec_meta}
Black-box meta-learning addresses the meta-learning problem \eqref{meta_general} by adopting a general-purpose optimizer for the per-period learning problem \eqref{opt_phi} as the adaptation procedure $\mathcal{A}(\cdot)$. Specifically, we adapt model agnostic meta-learning (MAML), a state-of-the-art meta-learning technique whose key idea is parametrizing the algorithm $\mathcal{A}(\cdot)$ with an initialization vector $\Phi_0 \in \R^{N \times L}$ used to tackle the inner problem \eqref{opt_phi} via SGD. In this section, we first develop MAML, as well as its simplified version, FOMAML, for power allocation via REGNNs. Then, we observe that black-box meta-learning does not affect the permutation equivariance of REGNNs highlighted in \cite{eisen2020optimal}.

\subsection{MAML and FOMAML}
MAML and FOMAML parametrize the adaptation algorithm with the initialization vector $\Phi_0$. Accordingly, assuming for simplicity a single step of gradient descent for problem \eqref{opt_phi}, we have the training algorithm
\begin{align}\label{locM}
    \mathcal{A}(\mathcal{D}^{\text{tr}}_\tau \, | \, \Phi_0) = \Phi_0 + \gamma \nabla_{\Phi_0} \left( \sum_{t \in \mathcal{D}^{\text{tr}}_\tau} \sum_{k = 1}^{K} c^k(G_\tau(t), \textrm{f} (G_\tau(t) \, | \, \Phi_0))\right),
\end{align}
where $\gamma > 0$ denotes the learning rate and we have made explicit the dependence on the initialization $\Phi_0$ in the notation $\mathcal{A}(\mathcal{D}^{\text{tr}}_\tau \, | \, \Phi_0)$. The update \eqref{locM} can be directly generalized to include multiple GD steps, as well as a reduced size of the mini-batch to implement SGD. Furthermore, the same update, and generalization thereof, apply also to the meta-test period $\tau_{\text{test}}$, yielding the model parameters $\Phi_{\tau_{\text{test}}} = \mathcal{A}(\mathcal{D}^{\text{tr}}_{\tau_{\text{test}}} \, | \, \Phi_0)$.

With definition \eqref{locM} of the training algorithm, MAML addresses the optimization problem \eqref{meta_general}, which is restated as the maximization
\begin{align}\label{meta_opt_phi}
    \underset{\Phi_0}{\text{max}} \,\,\,\, & \sum_{\tau = 1}^{\mathcal{T}^{\text{meta}}} \sum_{k = 1}^{K_\tau} \sum_{t \in \mathcal{D}^{\text{te}}_\tau} c^k(G_\tau (t), \textrm{f} (G_\tau (t) \, | \, \Phi_\tau = \mathcal{A}(\mathcal{D}^{\text{tr}}_\tau \, | \, \Phi_0))),
\end{align}
over the initialization $\Phi_0$.

For the single GD update in \eqref{locM}, the meta-training problem in \eqref{meta_opt_phi} is addressed by MAML using GD, which updates the initialization $\Phi_0$ in the outer loop as
\begin{align}\label{locMsecond}
    \Phi_0 \leftarrow &\Phi_0 
    - \delta \left( \I - \sum_{\tau = 1}^{\mathcal{T}^{\text{meta}}} \sum_{k = 1}^{K_\tau} \sum_{t \in \mathcal{D}^{\text{te}}_\tau} \nabla^2_{\Phi_0} c^k(G_\tau(t), \textrm{f} (G_\tau(t) \, | \, \Phi_0)) \right) 
    \nonumber \\&
    \times \left( \nabla_{\Phi_\tau} \sum_{\tau = 1}^{\mathcal{T}^{\text{meta}}}\sum_{k = 1}^{K_\tau}  \sum_{t \in \mathcal{D}^{\text{te}}_\tau} c^k(G_\tau(t), \textrm{f} (G_\tau(t) \, | \, \Phi_\tau = \mathcal{A}(\mathcal{D}^{\text{tr}}_\tau \, | \, \Phi_0) )) \right),
\end{align}
where $\I$ denotes the identity matrix and $\delta > 0$ denotes the learning rate. Extensions to SGD are straightforward. 

The MAML update in \eqref{locMsecond}, requires computation of the Hessian of the REGNN mapping \eqref{regnn2} with respect of the model parameters, which can be expensive. First-order methods, such as FOMAML \cite{finn2017model}, aim at circumventing the need for computation of higher-order derivatives.
In particular, FOMAML ignores the Hessian terms in the updates of the shared parameters in \eqref{locMsecond}, obtaining the update
\begin{align}
    \Phi_0 &\leftarrow \Phi_0 
    + \delta \nabla_{\Phi_\tau} \left( \sum_{\tau = 1}^{\mathcal{T}^{\text{meta}}} \sum_{k = 1}^{K_\tau} \sum_{t \in \mathcal{D}^{\text{te}}_\tau}  c^k(G_\tau(t), \textrm{f} (G_\tau(t) \, | \, \Phi_\tau = \mathcal{A}(\mathcal{D}^{\text{tr}}_\tau \, | \, \Phi_0))) \right).
\end{align}
Algorithm 1 provides a summary of FOMAML for power allocation. The algorithm has a nested loop structure, with the outer loop updating the shared initialization parameters $\Phi_0$ and the inner loop carrying out the local model updates in \eqref{locM}. 

\subsection{Permutation Equivariance and Invariance}
An important property of REGNNs is their equivariance to permutations \cite{sandryhaila2013discrete}. {\color{black}{In the context of wireless networks, the equivariance and invariance properties imply that a relabelling or reordering of the transmitters in the network produces a corresponding permutation of the power allocation vector without any permutation of the filter weights. This essential structural property is not satisfied by general fully connected models, in which a restructuring of the network would require an equivalent permutation of the inter-layer weights. The outlined properties of REGNNs stem from the locality of the operations implemented in a GNN: the power used by a node depends only on information at a distance of $n$ hops in the interference graph, rather than global information about the entire network -- a practical requirement for a power control policy. }} In this subsection, we briefly review this important property, and observe that the solution provided by black-box meta-learning is also permutation invariant. 

Formally, let $\Pi$ denote a $K \times K$ permutation matrix such that the product $\Pi^T x$ reorders the entries of any given $K \times 1$ vector $x$, and the product $\Pi^T G \Pi$ reorders the rows and columns of any given $K \times K$ matrix $G$.
The output of the REGNN $\textrm{f} (G \, | \, \Phi)$ is permutation equivariant in the sense that, for a permutation matrix $\Pi$ and  channel matrix $G$, we have
\begin{align}\label{eqivBB}
    \textrm{f} (\Pi^T G \Pi \, | \, \Phi) = \Pi^T \textrm{f} (G \, | \, \Phi).
\end{align}

By \eqref{eqivBB}, the meta-learning objective in \eqref{meta_opt_phi} is permutation invariant in the sense that, for a permutation matrix $\Pi$ and any realizations of the channel matrices $G_\tau(t)$, we have
\begin{align}\label{eqiv_sum_rate}
    \sum_{\tau = 1}^{\mathcal{T}^{\text{meta}}} \sum_{k = 1}^{K_\tau} \sum_{t \in \mathcal{D}^{\text{te}}_\tau} c^k(\hat{G}_\tau (t), \textrm{f} (\hat{G}_\tau (t) \, | \, \Phi_\tau)) = \sum_{\tau = 1}^{\mathcal{T}^{\text{meta}}} \sum_{k = 1}^{K_\tau} \sum_{t \in \mathcal{D}^{\text{te}}_\tau} c^k(G_\tau (t), \textrm{f} (G_\tau (t) \, | \, \Phi_\tau)),
\end{align}
where $\hat{G}_\tau(t) = \Pi^T G_\tau(t) \Pi$.
As a consequence of the invariance of the objective in \eqref{eqiv_sum_rate}, the initialization produced by MAML in \eqref{locMsecond} is also invariant to permutations. 

\begin{algorithm}
\caption{Power allocation via black-box meta-learning (FOMAML)}

\begin{algorithmic}[1]

\Procedure{Offline meta-training}{}       
    \vspace{1mm}
    \State Initialize filter taps $\Phi_0$
    \vspace{1mm}
    \For{$I \,\,\, \text{meta-training iterations}$}
    \vspace{1mm}
        \State Select meta-training periods $\tau \in \{1,...,\mathcal{T}^{\text{meta}}\}$
        \vspace{1mm}
        \For{$\text{all selected meta-training periods} \,\,\, \tau$}
        \vspace{1mm}
        \State Update filter taps $\Phi_\tau = \mathcal{A}(\mathcal{D}^{\text{tr}}_\tau \, | \, \Phi_0)$ using data set $\mathcal{D}^{\text{tr}}_\tau$ according to (11)
        \vspace{1mm}
        \EndFor
        \vspace{1mm}
        \State Update shared initialization $\Phi_{0}$ using data set $\mathcal{D}^{\text{te}}_\tau$ according to (14)
        \vspace{1mm}
    \EndFor
\vspace{1mm}
\EndProcedure
\Return {shared initialization $\Phi_{0}$}
\vspace{1mm}
\Procedure{Adaptation at runtime}{} 
\vspace{1mm}
\For{$I \,\,\, \text{meta-testing iterations}$}
\vspace{1mm}
        \State Update filter taps $\Phi_{\tau_{\text{test}}} = \mathcal{A}(\mathcal{D}^{\text{tr}}_{\tau_{\text{test}}} \, | \, \Phi_0)$ using data set $\mathcal{D}^{\text{tr}}_{\tau_\text{test}}$ according to (11)
    \EndFor
\EndProcedure

\end{algorithmic}
\end{algorithm}


\section{Modular Meta-Learning}\label{sec_meta_mod}
The black-box meta-learning method described in the previous section aims at fast parametric generalization, sharing an initialization of the model parameters across periods. In this section, we propose a modular approach that aims at combinatorial generalization, finding a set of reusable modules that can form components of a solution for a new period. The distinction between the two approaches is illustrated in Fig~\ref{ML_comparison}. As seen in the figure, in modular meta-learning, the adaptation algorithm $\mathcal{A}(\cdot)$ selects the filters $\Phi_l$ to be applied at each layer $l=1,...,L$ of the REGNN \eqref{regnn2} from a shared module set $\mathcal{M} = \{\Phi^{(1)},...,\Phi^{(M)}\}$, representing a repository of filter taps. The key idea is that the module set $\mathcal{M}$ is optimized during meta-training, while it is fixed at runtime, enabling an efficient adaptation based on limited data via the selection of modules from $\mathcal{M}$. {\color{black}{Existing modular meta-learning methods \cite{alet2018modular} rely on global optimization methods based on simulated annealing to find suitable combinations of modules. Accordingly, the module assignment procedure is fixed, and only the modules are designed via meta-learning. In contrast, the proposed method adopts a stochastic module assignment, which enables the efficient joint optimization of modules and module assignment procedure via the Gumbel-softmax reparametrization trick \cite{maddison2016concrete} and standard SGD. Through the proposed approach, module assignment does not require the application of a global optimization procedure, but rather the efficient application of local SGD updates. }}

\subsection{Modular Meta-learning}
A module assignment $S_\tau \in \{1,...,M\}^L$ is a mapping between the layers $l = 1,...,L$ of the REGNN and the modules from the module set $\mathcal{M}$. Mathematically, the assignment $S_\tau = \left[S_{\tau, 1},...,S_{\tau, L}\right]^T$ is an $L$-dimensional vector, with the $l$-th element $S_{\tau, l} = [S_\tau]_l \in \{1,...,M\}$ indicating the module assigned to layer $l$ at period $\tau$. Thereby, the assignment vector $S_\tau$ can take $M^L$ possible values. Let us represent the categorical variable $S_{\tau, l}$ using a one-hot representation $S_{\tau, l} = [S^{(1)}_{\tau, l},...,S^{(M)}_{\tau, l}]^T$, in which $S^{(i)}_{\tau, l}=1$ if $S_{\tau, l} = i$, and $S^{(i)}_{\tau, l}=0$ otherwise. With this definition, we can write the output \eqref{regnn1} of layer $l$ of the modular REGNN as
\begin{align}\label{mod_regnn}
    z_{l+1} = \sum_{i = 1}^{M} S_{\tau, l}^{(i)} \left[ \sigma \left[ \sum_{n=1}^N \Phi^{(i)}_{n} (G_\tau (t))^n z_l \right]\right].
\end{align}
Using a recursive application of \eqref{mod_regnn}, for a given module set $\mathcal{M}$ and module assignment vector $S_{\tau}$, the transmit power can be found as the output of the modular REGNN as
\begin{align}\label{regnn2_mod}
    &\textrm{f} \left(G_\tau(t) \, | \, \left[\Phi^{(S_{\tau, 1})},...,\Phi^{(S_{\tau, L})}\right]\right) = P_\text{\textrm{max}} \nonumber\\&
    \times \sum_{i = 1}^{M} S_{\tau, L}^{(i)} \left[ \sigma \left[ \sum_{n=1}^N \Phi^{(i)}_{n} (G_\tau (t))^n \left( ... \left(\sum_{i = 1}^{M} S_{\tau, 1}^{(i)} \left[ \sigma \left[ \sum_{n=1}^N \Phi^{(i)}_{n} (G_\tau (t))^n x \right]\right) ...\right) \right] \right] \right].
\end{align}

The objective during meta-training is to optimize a module set $\mathcal{M}$ that allows the system to find a combination of effective modules for any new topology during deployment. This is done by formulating problem \eqref{meta_general} as the maximization
\begin{align}\label{opt_phi_meta}
    \underset{\mathcal{M}}{\text{max}} \,\,\,\, &\sum_{\tau = 1}^{\mathcal{T}^{\text{meta}}} \sum_{k = 1}^{K_\tau} \sum_{t \in \mathcal{D}^{\text{te}}_\tau} c^k(G_\tau (t), \textrm{f} (G_\tau (t) \, | \, \Phi_\tau = \mathcal{A}(\mathcal{D}^{\text{tr}}_\tau \, | \, \mathcal{M})))
\end{align}
over the module set $\mathcal{M}$, where the learning algorithm $\mathcal{A}(\mathcal{D}^{\text{tr}}_\tau \, | \, \mathcal{M})$ selects the best possible assignment from set $\mathcal{M}$ given CSI data $\mathcal{D}^{\text{tr}}_\tau$. Accordingly, the training algorithm is given as a function of the module set $\mathcal{M}$ as
\begin{align}\label{opt_phi_A}
    \mathcal{A}(\mathcal{D}^{\text{tr}}_\tau \, | \, \mathcal{M}) = \left[\Phi^{(S^*_{\tau, 1} (\mathcal{M}))},...,\Phi^{(S^*_{\tau, L}(\mathcal{M}))}\right],
\end{align}
where the optimized assignment vector is
\begin{align}\label{opt_s}
    S_\tau^*(\mathcal{M}) = \underset{S_\tau \in \{1,...,M\}^L}{\text{argmax}} \sum_{k = 1}^{K_\tau} \sum_{t \in \mathcal{D}^{\text{tr}}_\tau}  c^k\left(G_\tau (t), \textrm{f} \left(G_\tau (t) \, | \, \left[\Phi^{(S_{\tau, 1})},...,\Phi^{(S_{\tau, L})}\right]\right)\right).
\end{align}

\subsection{Determining the Module Assignment}
The optimization \eqref{opt_phi_meta} is a mixed continuous-discrete problem over the module set and the assignment variables $\{S_\tau\}_{\tau = 1}^{\mathcal{T}^{\text{meta}}}$. To address this challenging problem, we define a stochastic module assignment function given by the conditional distribution $\mathcal{P}_\tau (S_\tau |  \mathcal{M}, \mathcal{D}^{\text{tr}}_\tau)$. This distribution assigns probabilities to each one of the $M^L$ possible assignment vectors $S_\tau$, given the module set $\mathcal{M}$ and training data $\mathcal{D}^{\text{tr}}_\tau$ for the current period $\tau$.
We can now redefine the bi-level optimization problem in \eqref{opt_phi_meta} as 
\begin{align}\label{opt_phi_meta_sto}
    &\underset{\mathcal{M}}{\text{max}} \,\,\,\, \sum_{\tau = 1}^{\mathcal{T}^{\text{meta}}} \underset{\mathcal{P}_\tau(\cdot \,| \, \mathcal{M}, \mathcal{D}_\tau^{\text{tr}})}{\text{max}} \mathbb{E}_{S_\tau \sim \mathcal{P}_\tau(S_\tau |  \mathcal{M}, \mathcal{D}^{\text{tr}}_\tau)} \left[ \sum_{k = 1}^{K_\tau} \sum_{t \in \mathcal{D}^{\text{te}}_\tau}  c^k(G_\tau (t), \textrm{f} (G_\tau (t) \, | \, \left[\Phi^{(S_{\tau, 1})},...,\Phi^{(S_{\tau, L})}\right])) \right],
\end{align}
where the inner optimization is over the distributions $\{\mathcal{P}_\tau(\cdot \,| \, \mathcal{M}, \mathcal{D}_\tau^{\text{tr}})\}_{\tau = 1}^{\mathcal{T}^{\text{meta}}}$. Problems \eqref{opt_phi_meta_sto} and \eqref{opt_phi_meta} are equivalent in the sense that they have the same solution. This is because the optimal distributions $\{\mathcal{P}_\tau(\cdot \,| \, \mathcal{M}, \mathcal{D}_\tau^{\text{tr}})\}_{\tau = 1}^{\mathcal{T}^{\text{meta}}}$ concentrate at the optimal module assignment vector \eqref{opt_s}. As detailed next, we propose to leverage the reparametrization trick to tackle the stochastic optimization in \eqref{opt_phi_meta_sto} via SGD.

To start, we model the module assignment distribution $\mathcal{P}_\tau(S_\tau |  \mathcal{M}, \mathcal{D}^{\text{tr}}_\tau)$ by using a mean-field factorization across the layers of the REGNN, i.e., 
\begin{align}\label{assignment_dist}
    \mathcal{P}_\tau(S_\tau \, | \, \mathcal{M}, \mathcal{D}^{\text{tr}}_\tau) = \prod_{l=1}^L \mathcal{P}_\tau(S_{\tau, l} \, | \, \mathcal{M}, \mathcal{D}^{\text{tr}}_\tau),
\end{align}
where $S_{\tau, l} \in \{1,...,M\}$ is the $l$-th entry of the vector $S_{\tau}$. This does not affect the equivalence of problems \eqref{opt_phi_meta} and \eqref{opt_phi_meta_sto} since the deterministic solution given by \eqref{opt_s} can be realized by \eqref{assignment_dist}.
Then, we let $\eta_{\tau, l} = [\eta^{(1)}_{\tau, l}, ..., \eta^{(M)}_{\tau, l}]^T$, be the vector of logits that parametrize the assignment probabilities through the softmax function as
\begin{align}
    \mathcal{P}_\tau(S_{\tau, l} = i \, | \, \mathcal{M}, \mathcal{D}^{\text{tr}}_\tau) = \frac{\exp(\eta_{\tau, l}^{(i)})}{\sum_{i'=1}^{M} \exp(\eta_{\tau, l}^{(i')})}.
\end{align}

The Gumbel-Max trick \cite{gumbel1954statistical}, \cite{hazan2012partition}, \cite{maddison2014sampling} provides a simple and efficient way to draw a sample $S_{\tau,l}$ from a categorical distribution with logits $\eta_{\tau, l}$ as
\begin{align}\label{GSsam_dis}
    {S}_{\tau, l}^{(i)} = \mathbbm{1} \,\,\, \left( \underset{i' \in \{1,...,M\}}{\text{argmax}} \,\,\, (\eta_{\tau, l}^{(i')} + \epsilon_l^{(i')}) = i \right), \,\,\,\, \text{for} \,\,\,\, i = 1,...,M,
\end{align}
where $\mathbbm{1} (x)$ denotes the indicator function which equals one if the assignment is true, and zero otherwise, {\color{black}{meaning that a specific module is assigned only if its respective noisy logit has the highest value}}; and  $\epsilon_l^{(i)}$ represent independent Gumbel variables obtained as
\begin{align}\label{epsilon}
    \epsilon_l^{(i)} = -\log (-\log (n^{(i)}_l)), \,\,\,\, \text{for} \,\,\,\, i = 1,...,M,
\end{align}
with $n^{(i)}_l$ being independent uniform random variables, i.e., $n^{(i)}_l \sim \text{Uniform} (0,1)$.
Thereby, using the Gumbel-Max trick \eqref{GSsam_dis}, the sampling of a discrete random variable is reduced to applying a deterministic function of the parameters $\eta_{\tau, l}^{(i)}$ to noise variables drawn from a fixed distribution. 

The argmax operation in \eqref{GSsam_dis} is not differentiable, making the optimization of the parameter vectors $\{\eta_{\tau, l}\}_{l=1}^L$ via SGD infeasible. To address this issue, references \cite{maddison2016concrete}, \cite{jang2016categorical} adopt the softmax function as a continuous, differentiable approximation. Samples from the resulting concrete distribution can be drawn according to
\begin{align}\label{GSsam}
    \tilde{S}_{\tau, l}^{(i)} = \frac{\exp((\eta_{\tau, l}^{(i)} + \epsilon_l^{(i)})/\lambda)}{\sum_{i=1}^{M} \exp((\eta_{\tau, l}^{(i)} + \epsilon_l^{(i)})/\lambda)}, \,\,\,\, \text{for} \,\,\,\, i = 1,...,M,
\end{align}
where the variables $\epsilon_l^{(i)}$ are drawn according to \eqref{epsilon}.
The temperature parameter $\lambda > 0$ controls the extent to which random variable $\tilde{S}_{\tau, l}^{(i)}$ resembles the one-hot representation \eqref{GSsam_dis}:
As the temperature tends to zero, the sample $\tilde{S}_{\tau, l}^{(i)}$ becomes identical to $S_{\tau, l}^{(i)}$. 


Regardless of the value of the temperature, substituting the distribution $\mathcal{P}_\tau(S_{\tau} \, | \, \mathcal{M}, \mathcal{D}^{\text{tr}}_\tau)$ with the distribution ${\tilde{\mathcal{P}}}_\tau(\tilde{S}_{\tau} \, | \, \mathcal{M}, \mathcal{D}^{\text{tr}}_\tau)$ in \eqref{opt_phi_meta_sto} allows us to address the inner optimization problems in \eqref{opt_phi_meta_sto} over the assignment probabilities. To this end, the objective in \eqref{opt_phi_meta_sto} is estimated by drawing samples $\epsilon_l^{(i)}$ from \eqref{epsilon} and plugging \eqref{GSsam} into the objective function in \eqref{opt_phi_meta_sto}. As a result, we obtain a differentiable function with respect to the parameters $\eta_{\tau, l}$, which can now be optimized via SGD. 

To elaborate, consider for simplicity a single sample $\{\{\epsilon_l^{(i)}\}_{l=1}^L\}_{i=1}^M$ of the Gumbel random variables in \eqref{epsilon}. For a fixed set $\mathcal{M}$, the inner optimization problem in \eqref{opt_phi_meta_sto} can be written as 
\begin{align}\label{opt_prob_pi}
    &\underset{\{\eta_{\tau,l}\}_{l=1}^L}{\text{max}} \left[ \sum_{k = 1}^{K_\tau} \sum_{t \in \mathcal{D}^{\text{te}}_\tau}  c^k(G_\tau (t), \textrm{f} (G_\tau (t) \, | \, \{\eta_{\tau,l}\}_{l=1}^L)) \right],
\end{align}
where we have defined 
\begin{align}\label{regnn3_mod}
    &\textrm{f} \left(G_\tau(t) \, | \, \{\eta_{\tau,l}\}_{l=1}^L \right) \nonumber\\& = P_\text{\textrm{max}}
    \times \sum_{i = 1}^{M} \frac{\exp((\eta_{\tau, L}^{(i)} + \epsilon_L^{(i)})/\lambda)}{\sum_{i=1}^{M} \exp((\eta_{\tau, L}^{(i)} + \epsilon_L^{(i)})/\lambda)}  \left[ \sigma \left[ \sum_{n=1}^N \Phi^{(i)}_{n} (G_\tau (t))^n \right.\right. \nonumber\\& \left. \left.  \left( ... \left(\sum_{i = 1}^{M} \frac{\exp((\eta_{\tau, 1}^{(i)} + \epsilon_1^{(i)})/\lambda)}{\sum_{i=1}^{M} \exp((\eta_{\tau, 1}^{(i)} + \epsilon_1^{(i)})/\lambda)} \left[ \sigma \left[ \sum_{n=1}^N \Phi^{(i)}_{n} (G_\tau (t))^n x \right]\right) ...\right) \right] \right] \right].
\end{align}
The gradient of \eqref{regnn3_mod} with respect to $\eta_{\tau, l}$ can be easily calculated to carry out the updates of the inner problem in \eqref{opt_phi_meta_sto}. For later reference, a single step of gradient descent, given the current module set $\mathcal{M}$ yields the update
\begin{align}\label{updates_pi}
    \eta^*_{\tau, l}(\mathcal{M}) = \eta_{\tau, l} - \gamma \sum_{k = 1}^{K_\tau} \sum_{t \in \mathcal{D}^{\text{te}}_\tau}  \nabla_{\eta_{\tau, l}} c^k(G_\tau (t), \textrm{f} (G_\tau (t) \, | \, \{\eta_{\tau,l}\}_{l=1}^L)),
\end{align}
where $\gamma > 0$ denotes the learning rate. 

Tackling the outer optimization problem in \eqref{opt_phi_meta_sto} is more challenging. Specifically, the optimal parameters of the assignment distribution, e.g. \eqref{updates_pi}, are a function of the module set, and hence updating set $\mathcal{M}$ also requires the partial derivative with respect to the module parameters of the optimized $\eta^*_{\tau, l}(\mathcal{M})$ for the inner maximization in \eqref{opt_phi_meta_sto}. 
However, in a manner similar to FOMAML (and other first-order black-box methods such as \cite{nichol2018firstorder}), we ignore the higher-order derivatives and update the parameters in the module set as
\begin{align}\label{updates_M}
    \mathcal{M} \leftarrow \mathcal{M} - \delta \sum_{\tau = 1}^{\mathcal{T}^{\text{meta}}} \sum_{k = 1}^{K_\tau} \sum_{t \in \mathcal{D}^{\text{te}}_\tau}   \nabla_{\mathcal{M}} c^k(G_\tau (t), \textrm{f} (G_\tau (t) \, | \, \{\eta_{\tau,l}(\mathcal{M})\}_{l=1}^L)),
\end{align}
where $\delta > 0$ denotes the learning rate and the gradient with respect to the module parameters is computed at the previous iterate $\mathcal{M}$. Using \eqref{updates_pi} and \eqref{updates_M}, we can address \eqref{opt_phi_meta_sto} by iterating over optimizing the assignment probability given the current module set, and optimizing the module parameters given the optimized assignment probability.

\subsection{Optimization During Runtime}
During meta-testing, we consider the obtained module set $\mathcal{M}$ as fixed. Using the training portion of the meta-test data set $\mathcal{D}^{\text{tr}}_{\tau_\text{test}}$, we only optimize the parameters of the distribution $\tilde{\mathcal{P}}_{\tau_\text{test}}(\tilde{S}_{\tau_\text{test}} | \mathcal{M}, \mathcal{D}^{\text{tr}}_{\tau_\text{test}})$ using \eqref{updates_pi}, or, more practically, multiple gradient descent steps. The final REGNN is constructed by using the mode of the assignment distribution as
\begin{align}\label{select_s}
    {S}_{\tau_\text{test}, l} = \underset{i \in \{1,...,M\}}{\text{argmax}} \,\,\, \tilde{\mathcal{P}}_{\tau_\text{test}}(\tilde{S}_{\tau_\text{test},l} = i \, | \, \mathcal{M}, \mathcal{D}^{\text{tr}}_{\tau_\text{test}}),
\end{align}
yielding the REGNN
\begin{align}\label{regnn2_adapt}
    &\textrm{f} \left(G_{\tau_\text{test}} \, | \, \left[\Phi^{(S_{\tau_\text{test}, 1})},...,\Phi^{(S_{\tau_\text{test}, L})}\right]\right) = P_\text{\textrm{max}} \nonumber\\&
    \times \left[ \sigma \left[ \sum_{n=1}^N \Phi^{(S_{\tau_\text{test}, L})}_{n} (G_{\tau_\text{test}} (t))^n \left( ... \left( \left[ \sigma \left[ \sum_{n=1}^N \Phi^{(S_{\tau_\text{test}, 1})}_{n} (G_{\tau_\text{test}} (t))^n x \right]\right) ...\right) \right] \right] \right].
\end{align}

Modular meta-learning is summarized in Algorithm 2.

\subsection{Permutation Equivariance and Invariance}
The modular nature of the REGNN in \eqref{regnn2_mod} does not violate the invariance properties of the individual filters, and of the module set by extension. 
To elaborate, observe that a single element in the assignment $S_\tau$ is non-zero, and, as a result, the output of the individual layers \eqref{mod_regnn} is equivalent to \eqref{regnn1}, whose equivariance properties have been established in \cite{eisen2020optimal}. Therefore, the composition in \eqref{regnn2_mod} is also equivariant,  as in \eqref{eqivBB},
and the objective in \eqref{opt_phi_meta} is invariant to permutation for any realization of the channel matrix $G(t)$ as in \eqref{eqiv_sum_rate}.
We conclude that the optimal module set $\mathcal{M}$ is invariant to permutations. 
In other words, any relabelling of the transmitters in the network will produce the same permutation of the power allocation without any modification of the taps in the module set. 

\begin{algorithm}
\caption{Power allocation via modular meta-learning}

\begin{algorithmic}[1]

\Procedure{Offline meta-training}{}       
    \vspace{1mm}
    \State Initialize module set $\mathcal{M}$
    \vspace{1mm}
    \For{$I \,\,\, \text{meta-training iterations}$}
    \vspace{1mm}
        \State Select meta-training periods $\tau \in \{1,...,\mathcal{T}^{\text{meta}}\}$
        \vspace{1mm}
        \For{$\text{all selected meta-training periods} \,\,\, \tau$}
        \vspace{1mm}
        \State Update the parameters of the assignment distribution $\tilde{\mathcal{P}}_\tau(\tilde{S}_\tau |  \mathcal{M}, \mathcal{D}^{\text{tr}}_\tau)$ using data set  $\mathcal{D}^{\text{tr}}_\tau$ according to \eqref{updates_pi}  
        \vspace{1mm}
        \EndFor
        \vspace{1mm}
        \For{$\text{all selected meta-training periods} \,\,\, \tau$}
        \vspace{1mm}
        \State Update module parameters $\mathcal{M}$ using data set $\mathcal{D}^{\text{te}}_\tau$ using \eqref{updates_M}
        \vspace{1mm}
        \EndFor
        \vspace{1mm}
    \EndFor
\vspace{1mm}
\EndProcedure
\Return {module set $\mathcal{M}$}
\vspace{1mm}
\Procedure{Adaptation at runtime}{} 
\vspace{1mm}
\For{$I \,\,\, \text{meta-testing iterations}$}
\vspace{1mm}
        \vspace{1mm}
        \State Update the parameters of the assignment distribution $\tilde{\mathcal{P}}_{\tau_\text{test}}(\tilde{S}_{\tau_\text{test}} |  \mathcal{M}, \mathcal{D}^{\text{tr}}_{\tau_\text{test}})$ using data set $\mathcal{D}^{\text{tr}}_{\tau_\text{test}}$  according to \eqref{updates_pi} 
        \vspace{1mm}
    \EndFor
    \vspace{1mm}
\State{Select module assignment using \eqref{select_s}  }  
\vspace{1mm}
\EndProcedure

\end{algorithmic}
\end{algorithm}

\section{Experiments}\label{sec_num}
In this section, we provide numerical results to elaborate on the advantages of black-box and modular meta-learning for power control in distributed wireless networks. 

\subsection{Network and Channel Model}
As in \cite{eisen2020optimal}, a random geometric graph in two dimensions comprised of $2 K_\tau$ nodes is drawn in each period $\tau$ by dropping each transmitter $k$ uniformly at random at location $\textrm{Tx}_{\tau,k} \in \left[-K_\tau, K_\tau \right]^2$, with its paired receiver $r^k_\tau$ at location $\textrm{Rx}_{\tau,k} \in \left[\textrm{Tx}_{\tau,k}-\frac{K_\tau}{4}, \textrm{Tx}_{\tau,k} + \frac{K_\tau}{4}\right]^2$. Given the geometric placement, the fading channel state between transmitter $k$ and receiver $j$ is given by 
\begin{align}
    h^{k,j}_\tau (t) = h^{k,j}_{\tau, \textrm{p}} (\cdot) h^{k,j}_{\tau, \textrm{f}} (t),
\end{align}
where the subscript $\textrm{p}$ denotes the path-loss gain which is invariant during a period $\tau$, and the subscript $\textrm{f}$ denotes the fast-fading component, which depend on the time slot $t$. The constant path-loss gain is given as $h^{k,j}_{\tau, \textrm{p}} = ||\textrm{Tx}_{\tau,k} - \textrm{Rx}_{\tau,k} ||^{-\gamma}$, where the path-loss exponent is set to $\gamma = 2.2$. The fast fading component $h^{k,j}_{\tau, \textrm{f}} (t)$ is random, and is drawn i.i.d. over indices $t$ and $\tau$ according to $|h^{k,j}_{\tau, \textrm{f}} (t)| \sim \text{Rayleigh} \, (1)$. Thereby, at each time slot $t$, fading conditions change, and the instantaneous channel information is used by the model to generate the optimal power allocation. The noise power is set to $\sigma^2 = -70$ dBm, and the maximum transmit power $P^k_{\text{max}}$ is set to $P^k_{\text{max}} = -35$ dBm for all devices. The corresponding maximum average SINR over the topology generation is
\begin{align}
    \frac{P^k_{\text{max}} \E \left[|h^{k,j}_{\tau, \textrm{p}}|^2\right]}{\sigma^2} &= \frac{P^k_{\text{max}} \E \left[ ||\textrm{Tx}_{\tau,k} - \textrm{Rx}_{\tau,k} ||^{-\gamma}\right]}{\sigma^2} \nonumber\\
    &\stackrel{(a)}{=} 35 \, \text{dB} + 10\log_{10} \left(\E [d^{-\gamma}]\right)\nonumber\\
    &\stackrel{(b)}{=} 35 \, \text{dB} + 10 \gamma \log_{10} \left(\frac{\sqrt{2} K_\tau}{8}\right),
\end{align}
where $d = ||\textrm{Tx}_{\tau,k} - \textrm{Rx}_{\tau,k} ||$ in $(a)$ is a uniform random variable, i.e., $d \sim \text{Uniform} \,\, \left(\left[0, \frac{\sqrt{2} K_\tau}{8}\right]\right)$, and $(b)$ follows from applying the Cavalieri's quadrature formula.
The large SNR implies that the system operates in the interference-limited regime, justifying the need for optimized power control policies. 
All details for the network and channel model are summarized in Table~\ref{table1} in Appendix B.

\subsection{Model Architecture and Hyperparameters}
As in \cite{naderializadeh2020wireless}, we consider a REGNN comprised of $L = 2$ hidden layers, each containing a filter of size $M = 4$. The non-linearity $\sigma [\cdot]$ in \eqref{regnn1} and \eqref{regnn2} is a ReLU, given by $\sigma[x] = \text{max}(0,x)$, except for the output layer where we use a sigmoid. Unless stated otherwise, the number of modules $M$ is set to $M=6$. In all experiments we set the input signal to an all-one vector. We define an annealing schedule for the temperature in \eqref{GSsam} over epochs, whereby the temperature is decreased in every epoch by $\exp{(-0.025)}$, until it reaches a predetermined minimal value, set to $\lambda_{\text{min}} = 0.5$ \cite{jang2016categorical}. 
All model hyper-parameters are summarized in Table~\ref{table2} in Appendix B.

\subsection{Data sets}
We study the case in which where the number of nodes in the network, $K_\tau$, is fixed, but the topology changes across periods; as well as the case in which the number of nodes in the network is also time-varying.

\subsubsection{Fixed network size}
In the first scenario, for a fixed number of links $K_\tau = 10$, each meta-training data set $\mathcal{D}_\tau$ corresponds to a realization of the random drop of the transmitter-receiver pairs at period $\tau$. Each drop is then run for $T_\tau = 100$ slots, whereby the fading coefficients are sampled i.i.d. at each slot. 
\subsubsection{Dynamic network size}
In the second scenario, the size of the network is chosen uniformly at random as $K_\tau \sim \text{Uniform} \,\, ([4, 20])$. Each meta-training data set $\mathcal{D}_\tau$ corresponds to a realization of the network size and to a random drop of the transmitter-receiver pairs as discussed above.

In both scenarios, unless stated otherwise, we set the number of meta-training periods to $\mathcal{T}^{\text{meta}} = 10$, and the training and the testing portions of the data set $\mathcal{D}_\tau$ contain $50$ slots each.
The meta-learning hyper-parameters are summarized in Table~\ref{table3} in Appendix B.

\subsection{Schemes and Benchmarks}
We compare the performance of the following schemes:
\subsubsection{Joint learning \cite{eisen2020optimal}} Adopted in \cite{eisen2020optimal}, joint learning pools together all tasks in the data set $\mathcal{D}^{\text{tr}}_\tau$, for all periods $\tau = 1,...,\mathcal{T}^{\text{meta}}$ in order to address problem \eqref{opt_phi} with an additional outer sum over periods $ 1,...,\mathcal{T}^{\text{meta}}$. The model parameters are then fine-tuned at runtime using the samples in the data set $\mathcal{D}_{\tau_\text{test}}$.

\subsubsection{Black-box meta-learning (Black-box ML)} As a representative black-box meta-learning method, we investigate the performance of FOMAML, as detailed in Algorithm 1. The number of gradient descent updates for both the task-specific and the shared parameters is set to $5$.

\subsubsection{Modular meta-learning (Modular ML)} We consider the proposed modular meta-learning method, as detailed in Algorithm 2. The number of gradient descent updates for the assignment and the module parameters are set to $5$.

{\color{black}{\subsubsection{Modular ML with exhaustive search} Since, as discussed, existing modular meta-learning schemes \cite{alet2018modular} rely on approximate global optimization, as a further benchmark, we consider modular meta-learning with exhaustive search to determine the module assignment. We note that the number of module combinations that need to be searched over is $M^L$, making it prohibitive to implement this strategy when the number of modules and layers is large. This reference scheme is meant to serve as an upper bound on the performance of the proposed scheme, which optimizes an efficient module assignment procedure based on SGD. The number of meta-training periods is set to $5$.}} 

\subsection{Results}
\begin{figure}[tbp]
\centering
\includegraphics[width=0.7\linewidth]{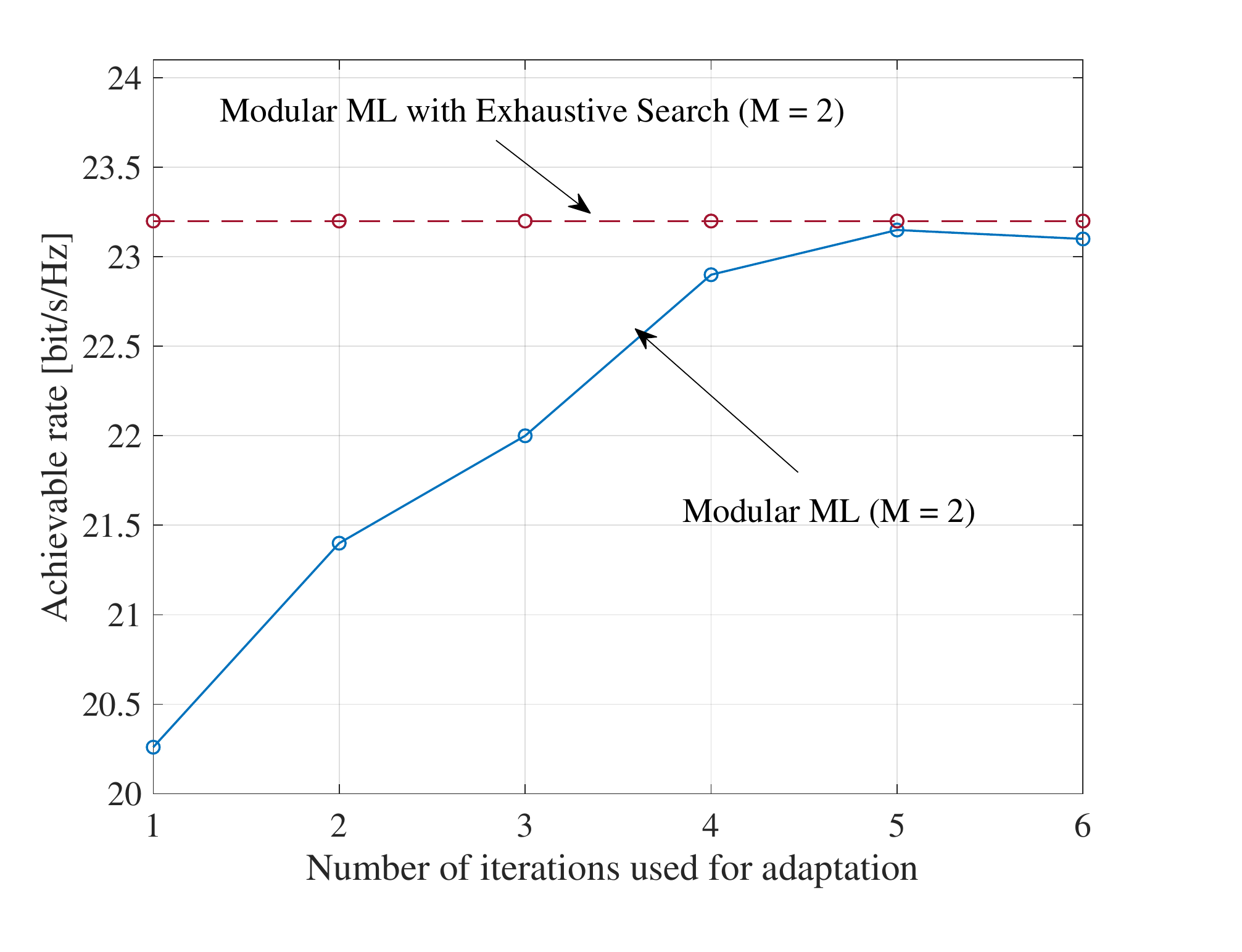}
\caption{Achievable sum-rate as a function of the number of iterations used for adaptation, comparing the performance of the proposed modular meta-learning scheme with that of an ideal method carrying out global module optimization. The number of training and testing samples for each task are set to $|\mathcal{D}^{\text{tr}}_\tau| = 50$, and $|\mathcal{D}^{\text{te}}_\tau| = 50$, respectively. The number of periods $\mathcal{T}$ is set to $\mathcal{T} = 5$. The results are averaged over $10$ independent trials.}
\label{rate_adaptation_iter}
\end{figure}

{\color{black}{\subsubsection{Comparison with exhaustive search for module assignment} To start, we present a toy example, in which we fix the number of modules to $M = 2$. The purpose of this experiment is to compare the proposed modular meta-learning method with an ideal, but far less efficient, solution in which the module assignment is determined by exhaustive search. This represents an ideal implementation of the scheme introduced in \cite{alet2018modular}, which relied on simulated annealing. In Fig.~\ref{rate_adaptation_iter}, we  investigate the achievable rates as a function of the number of iterations used for adaptation. The rate achieved with exhaustive search is depicted as a fixed value independent of the number of iterations, since this reference approach carries out  adaptation by looping over all $M^L = 16$ possible module combinations to determine the best combination. The figure shows that, with only $5$ iterations used for adaptation, the proposed stochastic approach can achieve equivalent rates, suggesting that the proposed efficient stochastic assignment can still result in a close-to-optimal selection of the modules. 

We note that the exhaustive-search benchmark is not further considered in the rest of this section due to its complexity for problems of practical size. Suboptimal global optimization schemes could be attempted, but the choice of a specific algorithm would entail additional and arbitrary design choices, such as defining suitable candidate solutions in the case of simulated annealing. }} 

\subsubsection{Runtime adaptation speed}
\begin{figure}[tbp]
\centering
\includegraphics[width=0.7\linewidth]{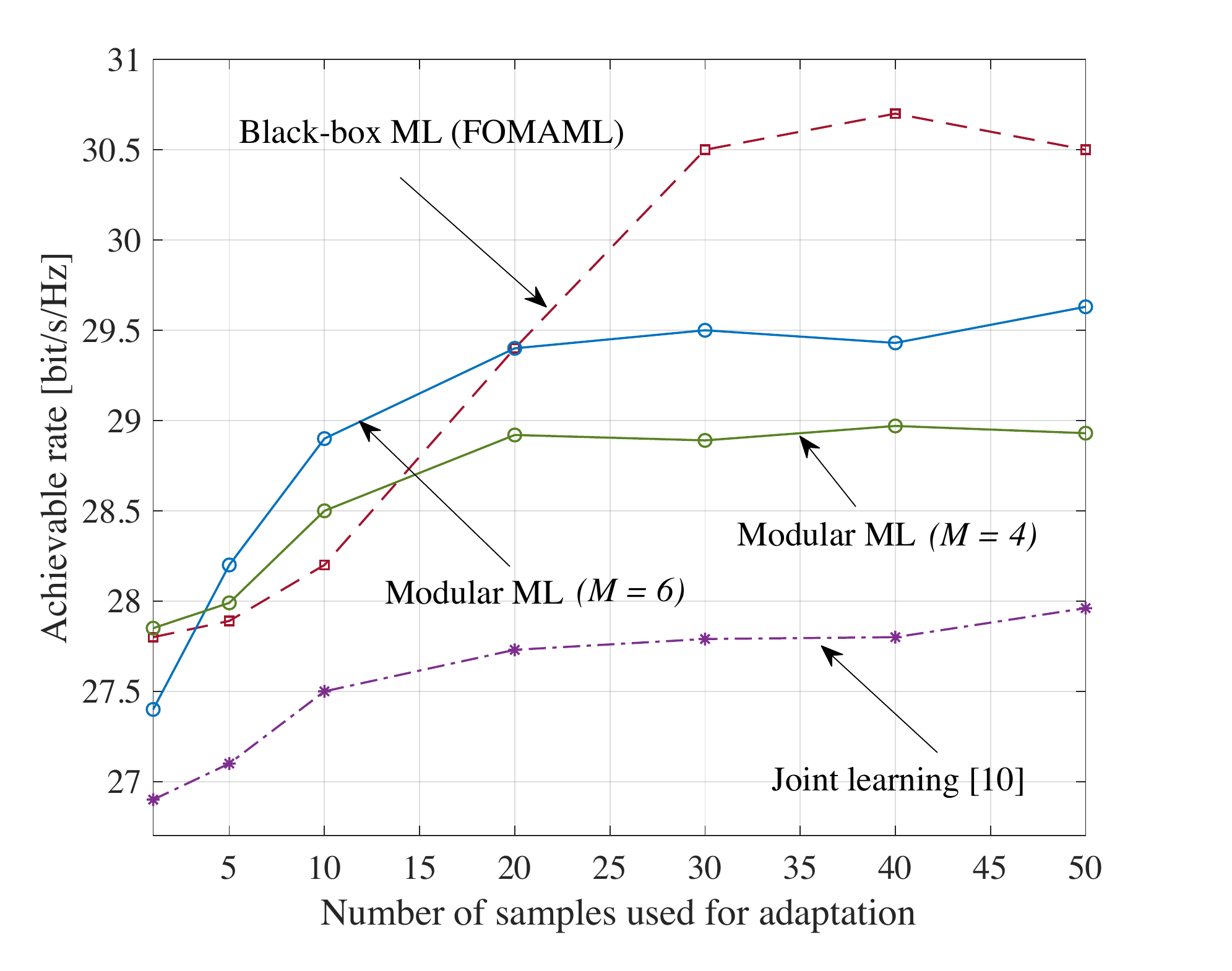}
\caption{Achievable sum-rate as a function of the number of samples used for adaptation. The number of training and testing samples for each task are set to $|\mathcal{D}^{\text{tr}}_\tau| = 50$, and $|\mathcal{D}^{\text{te}}_\tau| = 50$, respectively. The number of periods $\mathcal{T}$ is set to $\mathcal{T} = 10$. The results are averaged over $10$ independent trials.}
\label{rate_adaptation}
\end{figure}

Next, we evaluate the requirements in terms of the number of samples in the data set $\mathcal{D}_{\tau_\text{test}}$ for the new, meta-test topology at runtime by plotting the sum-rate as a function of the size of the data set $\mathcal{D}_{\tau_\text{test}}$ in Fig.~\ref{rate_adaptation}. We consider the more challenging case of networks with dynamic size. Fig.~\ref{rate_adaptation} confirms that meta-learning can adapt quickly to a new topology, using a much reduced number of samples, as compared to joint learning \cite{eisen2020optimal}. This validates the application of meta-learning to challenging communication problems like power control. Furthermore, modular meta-learning with both $M = 4$ and $M = 6$ modules is observed to outperform black-box methods when few adaptation samples are available, with the caviat that, a single adaptation sample is insufficient to determine a suitable module assignment when the number of modules is sufficiently large (here $M = 6$). This points to the benefits of a stronger (meta-)inductive bias in the regime where data availability is very limited. In particular, in modular ML, the adaptation samples are only used determine the module assignment at runtime and not to optimize the module parameters. As the number of samples for adaptation increases, the number of required modules grows and eventually black-box ML becomes advantageous. Overall, the results in Fig.~\ref{rate_adaptation} reveal a tension between the sample efficiency of modular ML and the flexibility of black-box methods.

\subsubsection{Offline data requirements}
\begin{figure}[tbp]
\centering
\includegraphics[width=0.7\linewidth]{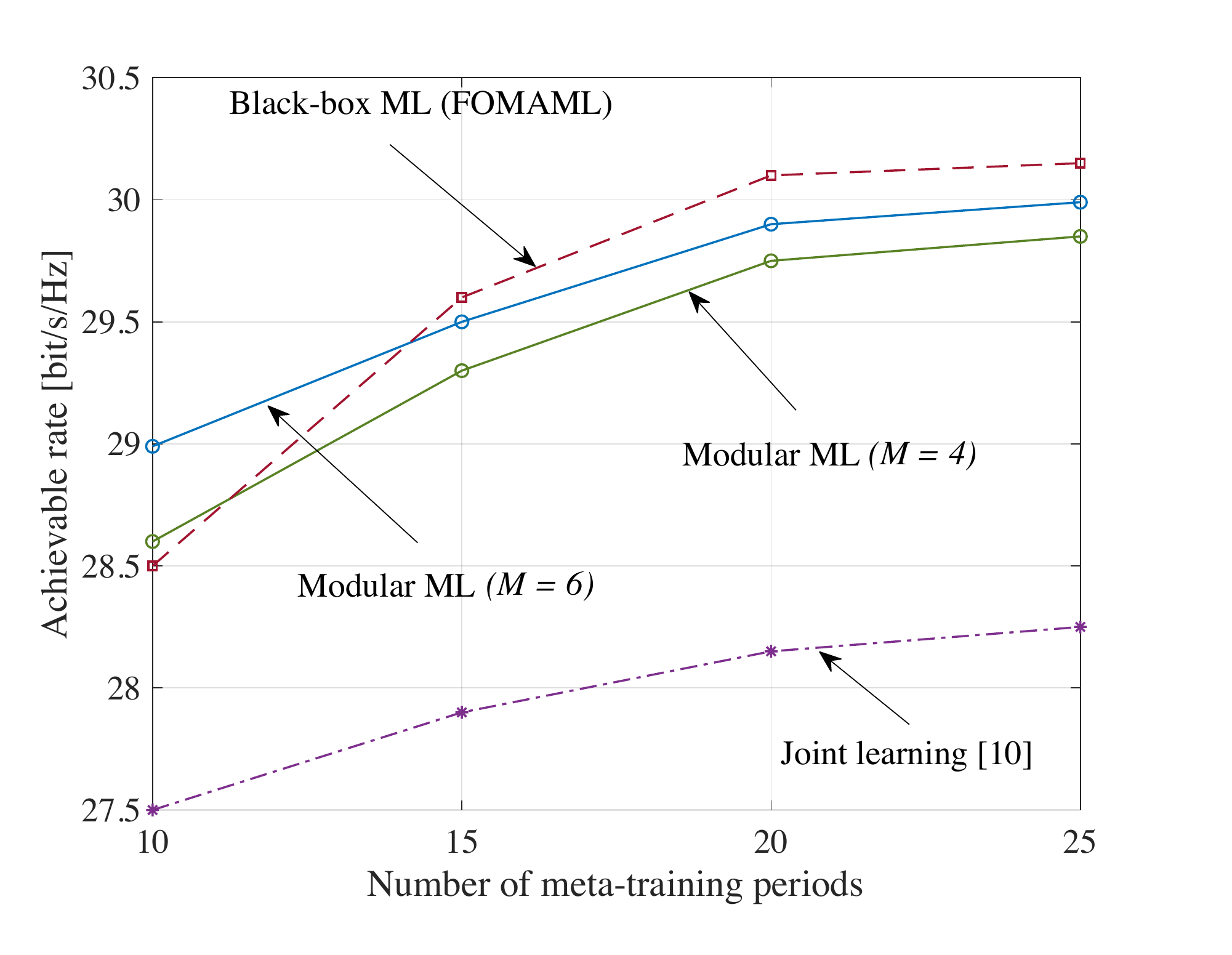}
\caption{Achievable sum rate as a function of the number of periods for meta-training. The number of training and testing samples for each task are set to $|\mathcal{D}^{\text{tr}}_\tau| = 50$, and $|\mathcal{D}^{\text{te}}_\tau| = 50$, respectively. The number of samples in the data set $\mathcal{D}_{\tau_\text{test}}$ is set to $10$. The results are averaged over $10$ independent trials.}
\label{rate_rept}
\end{figure}

We move to investigating the effect of the number $\tau_{\text{meta}}$ of periods, observed in the offline phase on the performance of joint learning and meta-learning for a network of dynamic size by plotting the sum-rate as a function of $\tau_{\text{meta}}$ in Fig.~\ref{rate_rept}. The results in Fig.~\ref{rate_rept} again demonstrate modular meta-learning to be advantageous over black-box methods when the number of meta-training tasks is smaller. However, due to the rigidity of modular methods, the gain is shown to be overcome by limitations due to bias as the number of meta-training tasks increases in which regime black-box methods are able to achieve larger rates.
In addition, the performance of all schemes saturates when there are sufficient periods for meta-training available.

\subsubsection{Impact of network variability}
\begin{figure}[tbp]
\centering
\includegraphics[width=0.7\linewidth]{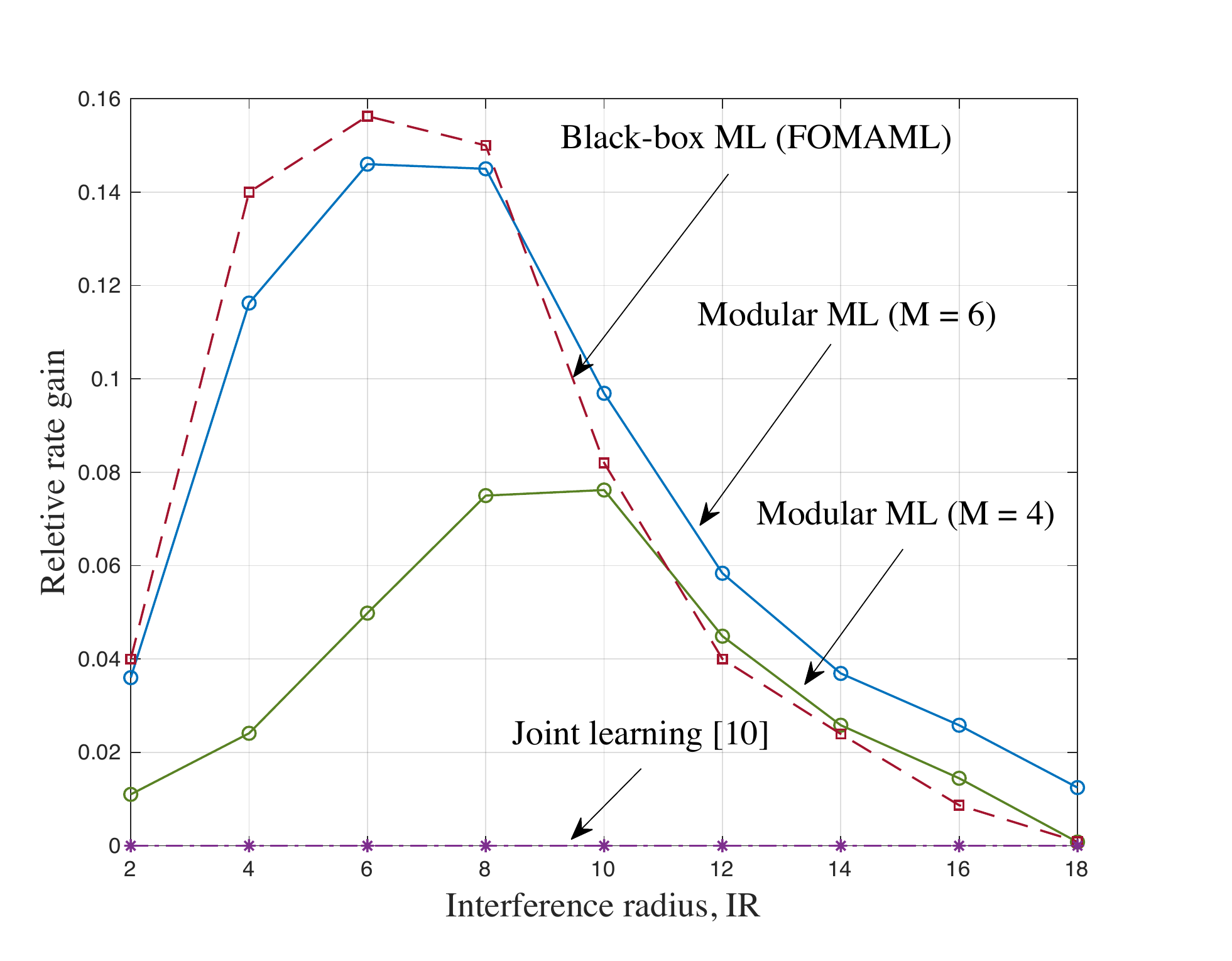}
\caption{Relative rate gain as a function of the interference radius. The number of training and testing samples for each task are set to $|\mathcal{D}^{\text{tr}}_\tau| = 50$, and $|\mathcal{D}^{\text{te}}_\tau| = 50$, respectively. The number of periods $\mathcal{T}$ is set to $\mathcal{T} = 10$. The number of samples in the data set $\mathcal{D}_{\tau_\text{test}}$ is set to $10$. The results are averaged over $10$ independent trials.}
\label{rel_rate_mod}
\end{figure}

To understand how topology changes affect the performance comparison between joint learning and meta-learning methods, we plot the relative achievable rate as a function of the interference radius (IR) in Fig.~\ref{rel_rate_mod} for a network of fixed size where $\mathcal{K}_\tau = 10$. The interference radius affects the size of the subset of interfering links, such that, depending on the location of the nodes, larger interference radius increases the number of interfering links. As a result, a small radius yields a fully disconnected graph at period $\tau$, while, as the interference radius increases, the graph becomes increasingly connected. At first, this produces a variety of topologies, until only a fully connected graph is obtained for sufficiently large values of the interference radius. Therefore, the distribution of the topologies is maximally diverse at some intermediate value of the interference radius. The relative rate gain is computed as $(C_{\text{ML}}-C_{\text{JL}})/C_{\text{ML}}$, where $C_{\text{ML}}$ and $C_{\text{JL}}$ are the sum rates obtained by the various meta-learning schemes and joint learning, respectively. 

Meta-learning is seen in Fig.~\ref{rel_rate_mod} to benefit from task diversity, achieving a large rate gain for intermediate values of the interference radius. Using only $M = 4$ modules in the set $\Phi$ is seen to be insufficient for modular meta-learning to capture the diversity in the topologies for some values of the interference radius as well as FOMAML does. Increasing the number of modules to $M = 6$ provides a larger rate gain, achieving a comparable performance to FOMAML, for all values of the interference radius.  However, the rate-gain dissipates for all meta-learning schemes as the task diversity decreases.

\subsubsection{Understanding modular ML}
\begin{figure*}[tbp]
\centering
\subfigure{\includegraphics[width=0.23\textwidth]{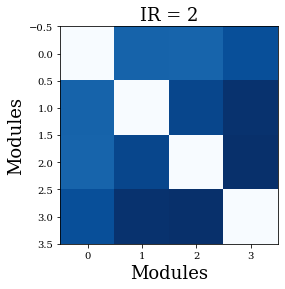}}
\subfigure{\includegraphics[width=0.23\textwidth]{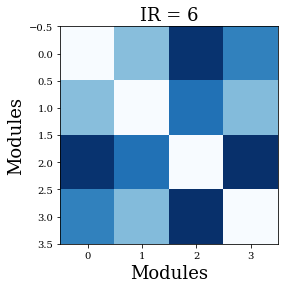}}
\subfigure{\includegraphics[width=0.23\textwidth]{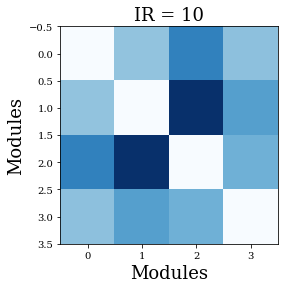}}
\subfigure{\includegraphics[width=0.28\textwidth]{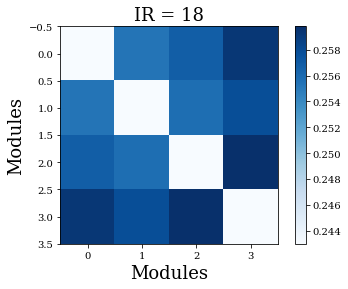}}
\caption{Linear CKA for different values of the interference radius (IR), set to $2$, $6$, $10$, and $18$. Darker shades represent higher CKA values.  The number of training and testing samples for each task are set to $|\mathcal{D}^{\text{tr}}_\tau| = 50$, and $|\mathcal{D}^{\text{te}}_\tau| = 50$, respectively. The number of periods $\mathcal{T}$ is set to $\mathcal{T}^{\text{meta}} = 10$. The number of samples in the data set $\mathcal{D}_{\tau_\text{test}}$ is set to $10$. The results are averaged over $10$ independent trials.}
\label{CKA}
\end{figure*}

In order to bring some insight into the operation of modular meta-learning, we now investigate the similarity between the modules in the set $\mathcal{M}$. As a similarity measure we adopt the linear centered kernel alignment (CKA) metric, first proposed in \cite{kornblith2019similarity}. Larger values of CKA indicate more similar modules (see Appendix A). Continuing the previous example, we vary the interference radius in order to focus on regimes with different variability across periods. Fig.~\ref{CKA} shows that large values of the interference radius IR result in similar modules and as a result high CKA values. On the other hand, intermediate values of the interference radius, which correspond to higher topological diversity, are shown to also induce a higher diversity in the modules. This variety in the module set in turn results in larger rates (see Fig.~\ref{rel_rate_mod}). 


\begin{figure*}[tbp]
\centering
\subfigure{\includegraphics[width=0.24\textwidth]{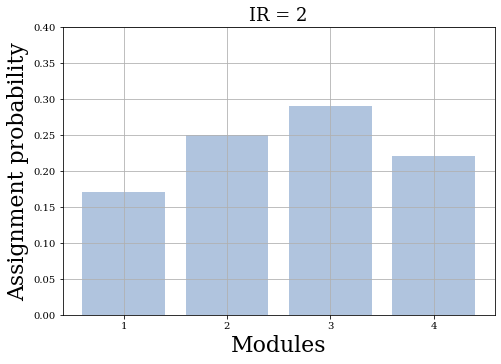}}
\subfigure{\includegraphics[width=0.24\textwidth]{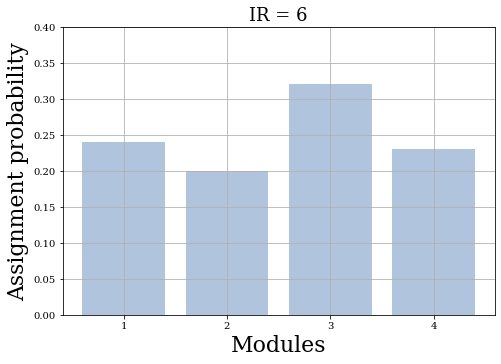}}
\subfigure{\includegraphics[width=0.24\textwidth]{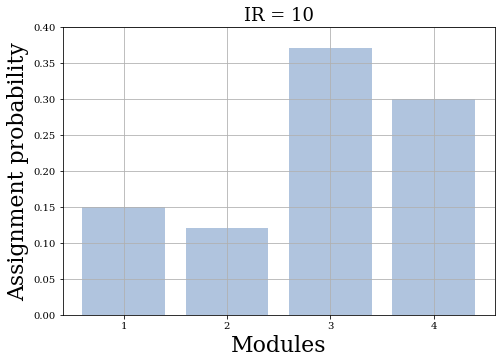}}
\subfigure{\includegraphics[width=0.24\textwidth]{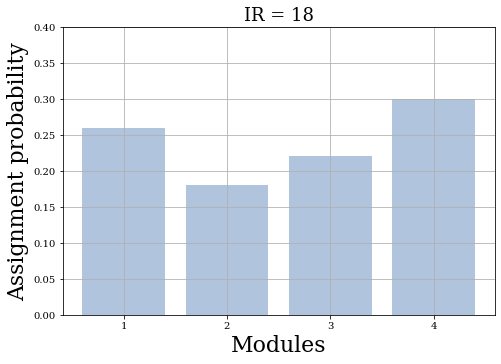}}
\caption{Assignment probability for different values of the interference radius (IR), set to $2$, $6$, $10$, and $18$. The number of training and testing samples for each task are set to $|\mathcal{D}^{\text{tr}}_\tau| = 50$, and $|\mathcal{D}^{\text{te}}_\tau| = 50$, respectively. The number of periods $\mathcal{T}$ is set to $\mathcal{T}^{\text{meta}} = 10$. The number of samples in the data set $\mathcal{D}_{\tau_\text{test}}$ is set to $10$. The results are averaged over $10$ independent trials.}
\label{select_prob}
\end{figure*}

To further elaborate on this observation, we present the assignment probability $\mathcal{P}_\tau(S_\tau \, | \, \mathcal{M}, \mathcal{D}^{\text{tr}}_\tau)$ for all modules in Fig.~\ref{select_prob} for different values of the interference radius. The assignment probability is estimated as the number of times a particular module is selected, averaged over all layers and trials. We first note that there are no idle modules, indicating an active participation from all of the modules for all values of the interference radius. Lower similarity between modules (see Fig.~\ref{CKA}), results in an uneven assignment distribution. For instance, module three is twice more likely to be selected than module two when the interference radius equals $10$. In contrast, when the similarity between modules is high, the assignment probability is shown to be more uniform. That is, for large values of the interference radius, there appears to be no clear preference for a particular module as the module set is more homogeneous.

\section{Discussion and Conclusion}\label{sec_con}
In decentralized wireless networks, meta-learning can enable quick adaptation of the power control policy to new network topologies by transferring knowledge from previously observed network configurations. This paper has investigated the integration of meta-learning and graph neural networks for power control problems, by proposing both black-box and modular meta-learning methods. An extensive experimental analysis has justified overall the use of meta-learning for power control in wireless networks. 

Notably, comparisons between the mentioned meta-learning schemes have revealed that modular meta-learning is preferable, outperforming black-box methods, in regimes requiring a stronger inductive bias, i.e., in regimes in which (very) limited meta-training data is available. This hints at other potential applications of modular meta-learning in communication engineering problems that are heavily constrained in terms of available data. For example, modular meta-learning may enable fast adaptation of channel access policies in IoT networks that are characterized by sporadic transmissions, significantly shortening the amount of time required for data acquisition.



\begin{appendices}
\section{Linear CKA}
Linear CKA \cite{kornblith2019similarity} is a similarity index that measures the relationship between functions. 
The similarity index between modules $i$ and $j$, $i \neq j$ can be found as
\begin{align}
    \text{CKA} (z^{(i)}, z^{(j)}) = \frac{|| z^{{(j)}^T} z^{(i)} ||_F}{|| z^{{(i)}^T} z^{(i)} ||_F || z^{{(j)}^T} z^{(j)} ||_F},
\end{align}
where $z^{(i)}$ and $z^{(j)}$ denote the outputs of individual modules $i$ and $j$, respectively, given by \eqref{mod_regnn}, and  $|| \cdot ||_F$ denotes the Frobenius norm. 

\section{Details of the experimental setup}
Details of the experimental setup are summerised in the tables below.


\begin{table*}[!h]
\centering
\caption{Network and Channel Model}
\label{table1}
\begin{tabular}{lll}
\hline
Hyper-parameters & Figs. 4,5,6  & Figs. 7,8,9         \\ \hline
Network size	       & $K_\tau \sim \text{Uniform} \,\, ([4, 20])$     & $K_\tau = 10$       \\ \hline
Path-loss exponent       & $\gamma = 2.2$      & $\gamma = 2.2$      \\ \hline
Noise power	   & $\sigma = -70$dBm	&$\sigma = -70$dBm	\\ \hline
Maximum power constraint	   & $P^k_{\text{max}} = -35$ dBm	& $P^k_{\text{max}} = -35$ dBm		\\ \hline
\end{tabular}
\end{table*}

\begin{table*}[!h]
\centering
\caption{Architecture hyper-parameters}
\label{table2}
\begin{tabular}{ll}
\hline
Hyper-parameters & Figs. 4,5,6,7,8,9      \\ \hline
Number of hidden layers & $2$   \\ \hline
Filter size & $N=4$  \\ \hline
Mini-batch size     & $64$   \\ \hline
Learning rate     & $0.0001$    \\ \hline
Optimizer     & Adam     \\ \hline
\end{tabular}
\end{table*}


\begin{table*}[!h]
\centering
\caption{Meta-learning hyper-parameters}
\label{table3}
\begin{tabular}{lllll}
\hline
Hyper-parameters & Fig. 4 & Fig. 5  & Fig. 6  & Figs. 7,8,9         \\ \hline
Number of meta-training periods	   & $\mathcal{T}^{\text{meta}} = 5$    & $\mathcal{T}^{\text{meta}} = 10$   & NA & $\mathcal{T}^{\text{meta}} = 10$     \\ \hline
Number of meta-training samples	       & $|\mathcal{D}^{\text{tr}}_\tau| = 50$   &
$|\mathcal{D}^{\text{tr}}_\tau| = 50$  & $|\mathcal{D}^{\text{tr}}_\tau| = 50$ & $|\mathcal{D}^{\text{tr}}_\tau| = 50$     \\ \hline
Number of meta-testing samples	       & $|\mathcal{D}^{\text{te}}_\tau| = 50$   &
$|\mathcal{D}^{\text{tr}}_\tau| = 50$   &
$|\mathcal{D}^{\text{te}}_\tau| = 50$ & $|\mathcal{D}^{\text{te}}_\tau| = 50$     \\ \hline
Number of adaptation samples   & $|\mathcal{D}_{\tau_{\text{meta}}}| = 10$  & NA   & $|\mathcal{D}_{\tau_{\text{meta}}}| = 10$ & $|\mathcal{D}_{\tau_{\text{meta}}}| = 10$     \\ \hline
Number of updates, task-spec. params., \\Black-box ML & NA  & $5$  & $5$ & $5$     \\ \hline
Number of updates, shared params., \\Black-box ML & NA  & $5$  & $5$ & $5$     \\ \hline
Number of updates, task-spec. params., \\Modular ML & NA  & $2$  & $2$ & $2$     \\ \hline
Number of updates, shared params., \\Modular ML  & $5$  & $5$  & $5$ & $5$     \\ \hline
Number of training updates, JL & NA  & $5$  & $5$ & $5$     \\ \hline
\end{tabular}
\end{table*}

\end{appendices}

\newpage

\bibliographystyle{IEEEtran}
\bibliography{litdab.bib}

\end{document}